\documentclass[12pt]{article}
\usepackage{mathptmx}
\usepackage[margin=2.5cm]{geometry}
\usepackage{graphicx}
\usepackage{subfigure}
\usepackage{eurosym}
\usepackage{verbatim}
\usepackage{microtype}
\usepackage{siunitx}
\usepackage{indentfirst}
\usepackage{algorithm}
\usepackage{amsmath}
\usepackage{algpseudocode}
\usepackage{url}
\newcommand*\rot{\rotatebox{90}}
\usepackage{apacite}
\let\cite\shortcite
\let\citeA\shortciteA
\usepackage[]{lineno}

\usepackage{titlesec}
\usepackage{authblk}
\usepackage{caption,threeparttable}
\titleformat*{\section}{\large\bfseries}
\titleformat*{\subsection}{\normalsize\bfseries}
\titleformat*{\subsubsection}{\normalsize\bfseries\itshape}

\title{\large \textbf{Ride-Pooling Matching with a Compensatory Cost Function: Implications for Adoption, Efficiency and Level of Service}}

\author[1,*]{\normalsize Arjan de Ruijter}
\affil{ Faculty of Civil Engineering and Geosciences, TU Delft}

\author[1]{Oded Cats}

\author[2]{Javier Alonso-Mora}
\affil{ Faculty of Mechanical, Maritime and Materials Engineering, TU Delft}

\author[1]{Serge Hoogendoorn}

\affil[*]{ Corresponding author: A.J.F.deRuijter@tudelft.nl - Stevinweg 1, Delft, Netherlands}

\date{\normalsize July 14, 2021}

\begin{document}
% \linenumbers
\maketitle
\begin{center}
\large
% \noindent \normalsize Submitted for review and possible publication in International Journal of Transportation Science and Technology \\
% \bigskip
% \noindent Corresponding author:\\
% Arjan de Ruijter \\
% Stevinweg 1 \\
% 2628 CN Delft, Netherlands \\
% A.J.F.deRuijter@tudelft.nl
\end{center}

\newpage

\section*{Abstract}
\noindent 
By utilising vehicle capacity more efficiently, ride-pooling platforms can potentially lead to reduced congestion levels without adversely prolonging travel times. While previous studies concluded that shared rides can offer substantial benefits, initial evidence suggests low adoption levels. We postulate that previous studies that investigated the potential of ride-pooling failed to account for the trade-off that users are likely to make when considering a shared ride. We address this shortcoming by formulating user net benefit stemming from sharing as a compensatory function where the additional travel time and on-board discomfort need to be compensated by the price discount for a traveller to choose a shared ride over a private ride. The proposed formulation is embedded in a method for matching travel requests and vehicles.
We conduct a series of experiments investigating how the potential of ride-pooling services depends on travel demand characteristics, user preferences and the pricing policy adopted by the service provider. In particular, the impact of various behavioural settings in terms of users’ willingness to share their ride and delay aversion on service adoption and its operational efficiency is assessed. Our results suggest that the total vehicle mileage savings found by previous studies is only attainable when users are very willing to share their ride (i.e. attach low premium to private rides) and are offered a 50\% discount for doing so. We find ride-pooling transportation distance savings as low as 15\% in less favourable behavioural scenarios.

\hfill\break%
\noindent\textit{Keywords}: Ride-pooling, Willingness to share, Delay tolerance, Demand distribution, Pricing
\newpage

\section{Introduction}
Recent developments in communication and information technologies have led to the rise of real-time and on-demand ride-pooling platforms like UberPool and ViaVan. Pre-pandemic, in New York alone on average more than 100,000 trips were booked daily using a ride-pooling application  \cite{todd2019nyc}. Users of those platforms consent to sharing a vehicle with travellers heading in a similar direction, even incurring small route deviations to pick-up and drop-off co-riders. By utilising vehicle capacity more efficiently, ride-pooling platforms can potentially lead to reduced congestion levels \cite{cici2013quantifying,wang2016pickup}.

The majority of the literature that quantified the societal benefits of ride-pooling concluded indeed that their introduction is expected to yield promising results \cite{wang2019ridesourcing}. A study by \citeA{Ma2013Tshare} for example stated that if the fleet of taxis in Beijing had allowed for shared rides in 2011, with users assumed to accept a maximum extra travel time of 5 minutes for their ride, 25\% more users could have been served and 13\% of the total vehicle distance could have been avoided. Another study analysed ride-pooling for up to three pooled requests using a graph representation that is denominated as a shareability graph and concluded that the total vehicle distance of taxis in New York could have been reduced by 32\% \cite{Santi2014quantifying}. Concurring evidence is offered by \citeA{qian2017optimal}, who show that with appropriate incentives total taxi vehicle mileage in New York can be reduced by 47\%, while a 46\% and 29\% mileage reduction is feasible in central areas of Wuhan and Shenzhen, respectively.
When ride-pooling is offered with high-capacity vehicles of up to ten seats, less than one fifth of the size of the current taxi fleet of New York City can serve 98\% of the original requests with a maximum delay of 3.5 minutes per passenger \cite{Alonso-Mora2017ondemand}. While the previously mentioned studies focused on ride-pooling in high-density metropolitan areas, \citeA{tachet2017scaling} assert that also cities with lower densities have the potential to obtain substantial efficiency gains by substituting individual taxi rides with ride-pooling.

Whether ride-pooling can live up to the potential indicated by the above studies, is highly uncertain. For example, the aforementioned studies have strictly analysed the existing demand for taxi services and assumed that ride-pooling services will substitute single-person taxi rides only. However, ride-pooling is also found to substitute other modes, like public transit and active modes \cite{rayle2016just}. This may lead to increased road traffic volumes. Moreover, as a result of low ride fares, the operation of a ride-pooling service may not be viable or require excessive subsidisation, and thus never fully materialise or scale. At the same time, the inconvenience associated with sharing a vehicle with strangers can be an important deterrent for potential users and therefore hinder a wide-scale adoption of ride-pooling services. In New York for example, prior to its shutdown in the wake of the COVID-19 pandemic, only 7\% of Uber's rides were made using its ride-pooling service UberPool \cite{todd2019nyc}. User inconvenience may arise from a lack of privacy \cite{Dueker1977ridesharing,Teal1987whohowwhy}, a feeling of dependence and a fear of having negative social interactions with other users \cite{Correia2011carpooling,morales2017share}. In the recent pandemic, virus exposure has emerged as an additional concern for shared mobility \cite{covid19maas}. Given that ride-pooling efficiency depends on mutual compatibility of trip requests, a low willingness to share is highly detrimental to the societal benefits of a ride-pooling system. In Toronto for example, only for 18\% of all ride-pooling trips that took place in September 2018 a rider was successfully matched to another rider \cite{toronto2019transportation}.

A possible explanation for the large discrepancy between the societal benefits of ride-pooling in theory and what has so far been observed in practice, is that previous ride-pooling studies fail to account for the complex trade-off that users are likely to make when considering a shared ride. In all of the aforementioned studies, users are assumed to opt for the ride-pooling service as long as the expected delay, i.e., later pick-up time and en-route detour, does not exceed a certain pre-specified threshold. Hence, no inherent deterrence from using ride-pooling was assumed given everything else being the same. In other words, in previous studies travellers were assumed to be intrinsically motivated to share a ride, without requiring any form of compensation and even accepting a delay.

The objective of this study is to explicitly account for the trade-off users can make between the discount offered for sharing their ride and the travel impedance that it induces. We formulate user net benefit due to sharing as a compensatory function where the additional travel time and on-board discomfort must be compensated by the shared ride discount for a traveller to choose a shared ride over a private one. This allows us to assess the potential adoption of ride-pooling by considering the two key barriers \cite{lavieri2019modeling}. In order to assess service performance and level of service, we embed our user benefit formulation in the method for matching travel requests and vehicles introduced by \citeA{Alonso-Mora2017ondemand}. 

The potential of ride-pooling services is expected to greatly vary across markets, depending on travel demand characteristics, user preferences and the pricing policy adopted by the service provider. Our approach enables that analysis of system performance under various settings while accounting for user preferences and their possible variation. We conduct a series of experiments that includes investigating the impact of various behavioural settings in terms of (a) users' willingness to share their ride and (b) their delay aversion on service adoption and its operational efficiency. The incorporation of a cost-benefit trade-off at the individual passenger level also allows us to outline implications for the design of an effective discount structure to boost ride-pooling adoption and consequently reduce the total vehicle distance on the road. 

The remainder of the paper is structured into four sections. Section \ref{sect:methodology} provides a detailed description of our methodology. This is followed by details on the design of the numerical experiment in Section \ref{sect:exp-design}. The results for the experiments are presented and discussed in Section \ref{sect:results}. The paper is concluded by stating the main conclusions that can be drawn in relation to the effect of users' behavioural preferences, the spatial distribution of demand and the pricing mechanism on the performance of a ride-pooling service (Section \ref{sect:discussion}).

\section{Methodology}\label{sect:methodology}
The assignment of passenger requests to ride-pooling vehicles over a period of time enables the assessment of the total vehicle movement and service quality obtained by a ride-pooling service. There are several approaches for the real-time assignment of requests to vehicles. As a way of dealing with the large solution space in ride-pooling assignment, in early assignment approaches, such as the one developed by \citeA{Ma2013Tshare}, incoming requests were individually allocated to vehicles using a greedy algorithm. \citeA{Santi2014quantifying} introduced the concept of shareability graphs to systematically analyse the mutual compatibility of two requests using a graph representation so that assignment can be performed with traditional graph-solving optimisation methods. A follow-up study extended this graph-based approach by introducing additional graph representations to allow for bundling requests and therefore compose high-occupancy ride-pooling trips \cite{Alonso-Mora2017ondemand}. Their request-group-vehicle (RGV) graph constitutes the composition of request groups and the vehicle that may serve each request group, representing the assignment problem as an Integer Linear Problem (ILP). A subsequent follow-up study by \citeA{simonetto2019real} further reduced the complexity of graph-based assignment approaches, without a significant loss of service performance. Agent-based models (ABM) have also been used to study ride-pooling, whereby users and vehicles are modelled as agents that dynamically interact \cite{fiedler2018impact,WINTER2018151}.

In this study we adopt the graph-based approach of \citeA{Alonso-Mora2017ondemand}, which offers the capability to model real-time ride-pooling with more than two passengers on-board the same vehicle. Contrary to a greedy approach, in which travellers are instantly assigned to an available vehicle when making a request, trip requests are being collected over a small period of time - in the order of seconds or minutes - and only assigned at the end of this interval. When given enough computation time, each assignment iteration will yield an optimal solution for the current set of requests, meaning that although the initial waiting time for travellers is longer, the approach can yield a shorter total travel time.

The approach contains a few implicit assumptions about ride-pooling operations. Firstly, supply is assumed to be centrally controlled: drivers are fully compliant with central instructions regarding which requests to serve, the order of pick-ups and drop-offs in a pooled ride, and the route to follow. Private and pooled services use the same road infrastructure, i.e. there are no HOV lanes for ride-pooling vehicles. To limit computational complexity in the approach, it is assumed that ride-pooling operations has no effect on the traffic conditions in the network. Travel times in the network are fully predictable and can thus be precomputed. On the user side, the approach assumes that requests can be subject to assignment in consecutive iterations. Assigned travellers that have not yet been picked-up can be reassigned if it improves their level of service. Travellers with unsatisfied requests will be available for assignment until a certain time threshold is reached, i.e. when they run out of patience.

Before providing more details on our methodological contribution to the ride-pooling assignment procedure, we will first shed more light on the graph-based assignment process of \citeA{Alonso-Mora2017ondemand}, which forms the basis for our work and is subject to adaptations as highlighted in the subsequent sections.

\begin{figure}[!ht]
  \centering
  \captionsetup{justification=centering}
  \includegraphics[width=0.75\textwidth]{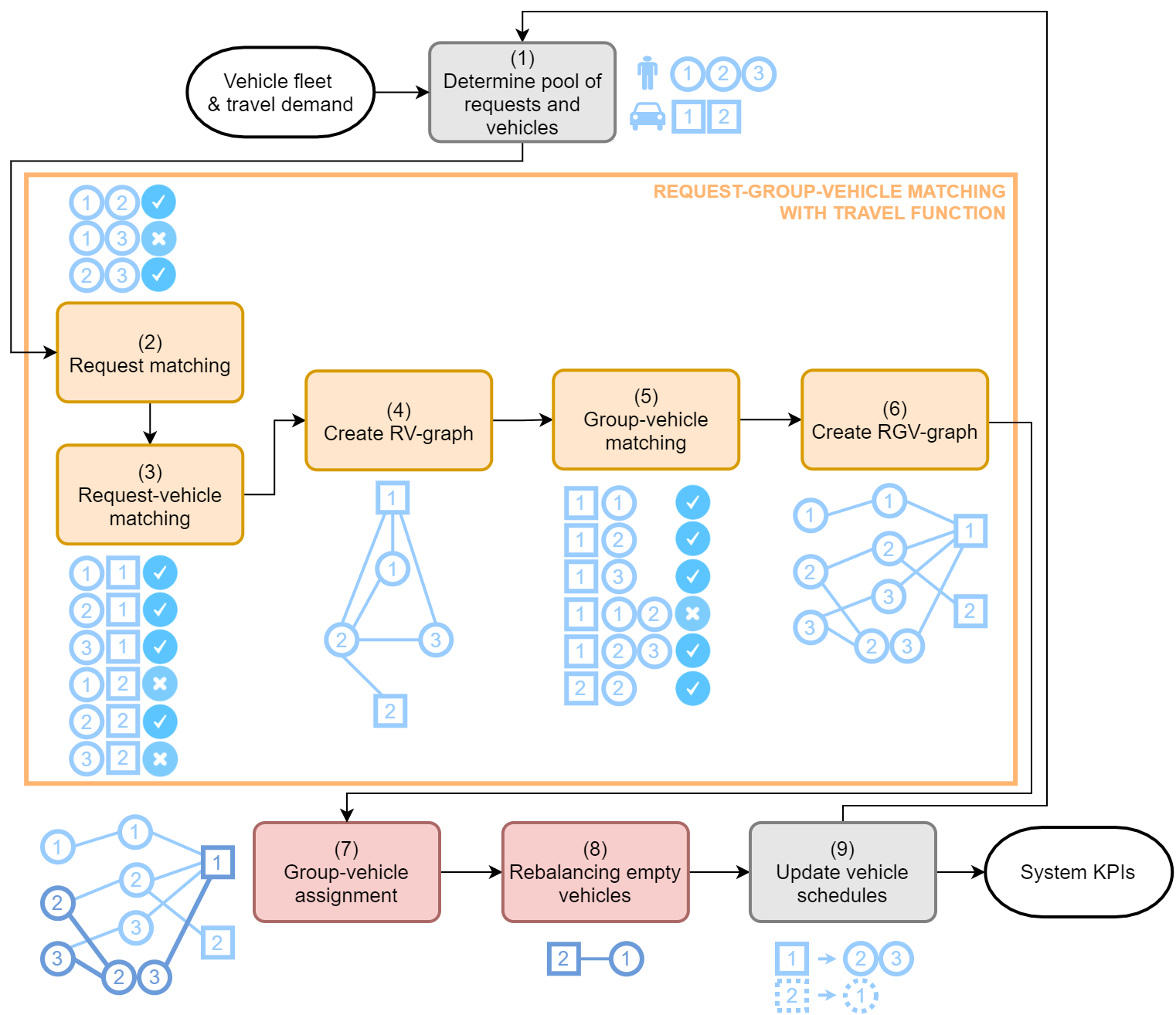}
  \caption{Overview of the methodology, including an example (in blue) with three requests and two (empty) vehicles.}\label{fig:Framework}
\end{figure}

\subsection{Framework}\label{subsect:frmw}
In the graph-based approach of \citeA{Alonso-Mora2017ondemand}, which is visualised in Fig. \ref{fig:Framework}, requests are matched with other requests as well as with vehicles at fixed intervals. Each assignment iteration consists of the following nine steps:

\begin{enumerate}
    \item Establish the status of vehicles as well as the pool of requests to be matched to a vehicle. The latter includes new trip requests and unassigned requests from the previous iteration. Pending requests are removed when assignment is no longer feasible. \\
    \textit{Illustration in Fig. \ref{fig:Framework}: two empty vehicles, three requests.}
    \item Check which pairs of requests can be matched, assuming there is a vehicle located at the origin of one of the two requests. \\
    \textit{Requests 1 and 2, and 2 and 3 form feasible request pairs.}
    \item Check which requests can be matched to which vehicles, considering vehicles' current location and on-board occupancy. \\
    \textit{All requests can be served by vehicle 1, only request 2 can be served by vehicle 2.}
    \item Create a request-vehicle (RV) graph: a graph with request and vehicle nodes, where an edge indicates a feasible request pair match (as found in step 2) or a feasible request-vehicle match (step 3). \\
    \textit{RV-graph as shown in Fig. \ref{fig:Framework}.}
    \item Check which groups of requests can be matched to which vehicles. To do so, first identify potentially feasible request group - vehicle matches based on cliques in the RV-graph, and then find whether there is a feasible route with which a vehicle can satisfy all requests in the group as well as passengers already on-board. \\
    \textit{From the RV-graph it follows that a group consisting of requests 1 and 2, and a group consisting of requests 2 and 3 can potentially be served by vehicle 1. Besides, all 'groups' with only one request can potentially be served by vehicle 1. Finally, a 'group' consisting only of request 2 can potentially be served with vehicle 2. When checking the feasibility of these group-vehicle combinations, we find that there is no feasible route for vehicle 1 to satisfy the group consisting of requests 1 and 2.}
    \item Create a request-group-vehicle (RGV) graph: a graph with request, group and vehicle nodes, where an edge between a request and a group node indicates that a request is included in a request group, and an edge between a group and a vehicle node implying that a request group can be matched to a vehicle (as found in step 5). A label can be added to edges between group and vehicle nodes to represent the attractiveness or 'value' of the match. \\
    \textit{RGV-graph as shown in Fig. \ref{fig:Framework}, with request (left), group (middle) and vehicle (right) nodes.}
    \item Decide which groups are assigned to which vehicles by translating the RGV-graph into an Integer Linear Problem. Multiple objective functions are possible, such as a maximisation of the number of assigned requests. \\
    \textit{The group consisting of requests 2 and 3 is assigned to vehicle 1. Vehicle 2 cannot serve request 1 and is left unassigned.}
    \item Decide whether idle vehicles will rebalance, and if so, where to. \\
    \textit{(Unassigned) vehicle 2 moves in the direction of (unassigned) request 1.}
    \item Update vehicle schedules according to the assignment (step 7) and rebalancing (step 8) result. \\
    \textit{Vehicle 1 will pick-up requests 2 and 3, vehicle 2 will rebalance towards the origin of request 1.}
\end{enumerate}

The next two sections provide a more detailed description of the method, focusing on the steps where this study differs from previous work. In the following, we present our version of the request-group-vehicle (RGV) matching procedure for finding all feasible request group - vehicle combinations (Subsection \ref{subsect:matching}, steps 2-6 in Fig. \ref{fig:Framework}), and the subsequent procedure of assigning vehicles optimally to requests (Subsection \ref{subsect:ass}, steps 7 and 8 in Fig. \ref{fig:Framework}).

\subsection{RGV-matching}\label{subsect:matching}
The matching procedure involves the creation of an RGV-graph to identify which group-vehicle combinations are feasible, given vehicle capacity and travel cost constraints. The matching process of \citeA{Alonso-Mora2017ondemand} is divided into three steps to ensure that not all group-vehicle combinations have to be enumerated and checked for feasibility, which significantly reduces the complexity of the matching algorithm. In this subsection, after we introduce the functionality of each of these three steps along with our modifications thereof, we provide a detailed formulation of the cost function and explain how it is used in the algorithm that checks the feasibility of group-vehicle pairs. 

The first amongst the matching steps (step 2 in Fig. \ref{fig:Framework}) establishes whether two requests in the pool of available requests $R$ can share a ride in the most fortunate scenario in which there is an empty vehicle readily available at the location of one of those requests. With this step, the set of potentially feasible request groups $G$ can already be significantly reduced. The next step (step 3 in Fig. \ref{fig:Framework}) checks whether a vehicle $v$ in the fleet of vehicles $V$ can serve a single request $r \in R$ given its current location and residual capacity (i.e. the number of available seats). The result of both steps can be combined and stored in a RV-graph (step 4) where edges indicate that two requests, or a request and a vehicle, can be matched. Each clique in the RV-graph represents a potentially feasible group-vehicle combination. Step 5 checks whether there exists a feasible sequence of stops $S_v$ for a vehicle $v$ to satisfy a group of requests $g\in G$, considering capacity and user level of service constraints. The RGV-graph consists of nodes representing the set of available requests $R$, the set of feasible request groups $G$ and the set of vehicles $V$. Each edge between a request $r \in R$ and a request group $g \in G$ has a label $a_{rg}$ indicating whether $r$ is included in $g$ ($a_{rg}=1$) or not ($a_{rg}=0$). Moreover, each edge connecting a request group $g \in G$ and a vehicle $v \in V$ has a label $b_{gv}$ indicating the sum of net sharing benefits of all requests in $g$ and passengers in $P_v$ given the optimal sequence of stops $S_v^*$. If a group-vehicle combination $gv$ is not feasible, $b_{gv}$ is assigned a very large penalty (a so-called big $M$), to ensure that this combination is not chosen during the assignment process. 

Our approach differs from the original one by \citeA{Alonso-Mora2017ondemand} in that group-vehicle edge labels in the RGV-graph contain sharing benefits instead of delay costs. Similarly, we substitute cost-based user constraints in the matching algorithm for benefit-based user constraints. In the remainder of this subsection we describe our modifications to the original work by \citeA{Alonso-Mora2017ondemand} in more detail.

\subsubsection{Net sharing benefit}
We formulate and quantify the individual benefits stemming from ride-pooling to represent the trade-off that travellers encounter when choosing between a private and a shared ride. In the following, we define the notion of net sharing benefits. It is used not only for identifying which requests and vehicles can be assigned to each other (i.e. in group generation and establishing group-vehicle pair feasibility), but also in deciding which requests are to eventually be assigned to which vehicles (i.e. in assignment), as explained in this and subsequent subsections.

We formulate the benefits and costs of ride-pooling relative to a private ride. The benefit associated with ride-pooling for a traveller with request $r$ corresponds to the total fare discount offered by the service provider for shared rides and is thus dependent on the discount rate $\pi_r$ that is applied to the ride fare $c_r$. $\pi_r$ can be either set as a fixed rate or as a function of the level of service offered by the shared ride. We examine both cases in our experiments.

In contrast, the disbenefits of a shared ride relative to a private ride relate to extra travel time imposed by sharing and an additional discomfort associated with sharing a vehicle, all other things being equal. The total disbenefit of a ride thus depends on how users perceive both attributes when expressed in monetary terms, in this study expressed as delay aversion $\beta_r$ and reluctance to share $\gamma_r$. These parameters indicate what fare discount users require for an hour of delay and for sharing a vehicle with other riders - value of time and ride-pooling alternative specific constant - respectively. To account for a different valuation of waiting time and in-vehicle delay, we introduce the parameter $\alpha_r$. It represents the additional user cost of waiting one minute for pick-up as opposed to having one minute of in-vehicle delay. The waiting time of a request $t_r^{wait}$ is calculated as the difference between the pick-up time $t_r^{pu}$ and time at which the request was made $t_r^r$. The total delay $t_r^{delay}$ is the difference between the actual time of drop-off $t_r^d$ and the earliest possible time of drop-off $t_r^* = t_r^r + tt_{o_r,d_r}$. The latter is calculated for an immediate pick-up and the shortest route, with travel time $tt_{o_r,d_r}$ between origin $o_r$ and destination $d_r$. 

The total net benefit of $r$ is thus formulated as:
\begin{alignat}{1}
    \qquad b_r = \pi_r \cdot c_r - (t_r^d - t_r^*) \cdot \beta_r - (t_r^{pu} - t_r^r) \cdot \alpha_r - \gamma_r \label{eq:benefit}
\end{alignat}

\subsubsection{Group-vehicle feasibility}
The three main matching steps in the RGV-approach (steps 2, 3 and 5 in Fig. \ref{fig:Framework}) involve applying the same algorithm for establishing whether a specific combination of requests and/or vehicles yields a feasible match. The steps differ only in their input: 

\begin{itemize}
    \item Pairwise request matching (step 2): two requests and a hypothetical vehicle located at the origin of one of the requests
    \item Request-vehicle matching (step 3): a single request and a specific vehicle
    \item Group-vehicle matching (step 5): a group of requests and a specific vehicle
\end{itemize}

The algorithm can therefore be generalised by stating that it finds whether a vehicle $v$ in fleet $V$, with passengers $P_v$ on-board, can serve a group of unassigned requests $U$. The vehicle can be either virtual or specific, and the group of requests may consist of a single request or of multiple requests.

In the first step of the algorithm, the complete set of stop sequences $K_v$ is identified with which $v$ can potentially satisfy $U$. Each stop sequence $S_v \in K_v$ is then checked for feasibility based on vehicle and user constraints. The vehicle constraint ensures that the vehicle capacity is not exceeded, i.e. $v$ cannot serve $U$ with stop sequence $S_v$ if the vehicle capacity $\kappa$ is exceeded by the vehicle occupancy $O_s$ after each stop $s\in S_v$, similar to the approach of \citeA{Alonso-Mora2017ondemand}. As we assume that drivers strictly follow the shortest path between two nodes in a network with static travel times, for each stop sequence $S_v$ we only need to check the feasibility constraints for a single route, i.e. the route made up of the shortest paths between the consecutive vehicle stops.

The user constraint proposed in this study is modelled with the net sharing benefit. Stop sequence $S_v$ is hereby assumed to satisfy the user level of service constraint only if the net benefit $b_r$ of each request in $U$ and $P_v$ is positive. In other words, all riders, whether already picked-up or not, must prefer a shared ride provisioned with this specific route over a private ride. 

If both types of constraints are satisfied, the benefit of the stop sequence $b_{S_v}$ is computed by summing the net benefits of all individual requests in $U$ and $P_v$. If there exists at least one feasible stop sequence $S_v\in K_v$ to serve $U$, then $U$ and $v$ form a feasible match. When there is more than a single feasible stop sequence, we will also need to determine which stop sequence is most optimal. Therefore, we search for the stop sequence in $K_v$ with maximum benefit $b_{S_v}$, which we define as the total benefit of the particular group-vehicle combination $b_{Uv}$. However, if no feasible stop sequence exists, $b_{Uv}$ is set to "Invalid". The complete procedure for checking the feasibility of group-vehicle combinations and finding the most optimal stop sequence for feasible combinations is specified in the pseudocode shown in Algorithm \ref{Alg:Route feasibility}.

\begin{algorithm}
\caption{Matching request group $U$ to vehicle $v$: Establishing feasibility and optimality of potential stop sequences.}\label{Alg:Route feasibility}
\begin{algorithmic}[1]
\footnotesize
\For {$S_v \in K_v$}
    \If {$\max_{s\in S_v} O_s \leq \kappa$} \Comment Vehicle capacity constraint \label{algline:ass-cap-constraint}
        \If {$b_r \geq 0, \forall r\in {U \cup P_v}$} \Comment User net benefit constraint \label{algline:ass-los-constraint}
            \State $b_{S_v} \gets \sum_{r\in U \cup P_v} b_r$ \Comment Total benefit of feasible stop sequence $S_v$ \label{algline:ass-tot-benefit}
        \Else
            \State ${b_{S_v}} \gets$ -1 \label{algline:ass-los-invalid}
        \EndIf
    \Else
        \State ${b_{S_v}} \gets$ -1 \label{algline:ass-cap-invalid}
    \EndIf
\EndFor
\If {$\exists S_v \in K_v$ such that $b_{S_v} > 0$}
    \State $b_{Uv} \gets \max_{{S_v}} b_{S_v}$  \Comment Stop sequence with maximum benefit \label{algline:ass-opt-route}
    \Else
    \State $b_{Uv} \gets "Invalid"$
\EndIf
\end{algorithmic}
\end{algorithm}

\subsection{Assignment \& rebalancing}\label{subsect:ass}
In this part of the method, requests are assigned to vehicles based on the RGV-graph (corresponding to step 7 in Fig. \ref{fig:Framework}). The group-vehicle assignment is treated as an Integer Linear Problem (ILP) with binary decision variables $x_{gv}$ indicating whether a group-vehicle combination with total net sharing benefit $b_{gv}$ is chosen or not. The ILP is defined as follows:

\begin{alignat}{4}
\max \quad & \sum_{g\in G} \sum_{v\in V} (b_{gv} + \sqrt{M} \cdot \sum_{r\in R} a_{rg}) \cdot x_{gv} \label{eq:ILP-ass-obj} \\
\textrm{s.t.} \quad & \sum_{g\in G} x_{gv} \leq 1, \; \forall v\in V \label{eq:ILP-ass-cstr-1} \\
& \sum_{g\in G} \sum_{v\in V} a_{rg} \cdot x_{gv} \leq 1, \; \forall r\in R \label{eq:ILP-ass-cstr-2} \\
& x_{gv} = [0,1], \; \forall g\in G, v\in V \label{eq:ILP-ass-cstr-3} 
\end{alignat}

\bigskip
The objective function (Equation \ref{eq:ILP-ass-obj}) aims at maximising the total benefit for accepted requests and passengers, while prioritising the acceptance of a maximum number of requests by adding a very large reward for each request $r$ in an assigned request group $g$. The sum of these rewards should however not overpass the big $M$ penalty assigned to infeasible group-vehicle combinations in the objective function. Therefore, the reward per request group is set to $\sqrt{M}$. The total benefit associated with a group-vehicle combination $g-v$ thus consists of the summed net benefit for all requests and passengers in this group plus a large reward $\sqrt{M}$ for each request that is a member of this group.

The Integer Linear Problem contains three types of constraints guaranteeing respectively a maximum assignment of one request group $g$ to each vehicle $v$ (Equation \ref{eq:ILP-ass-cstr-1}), that each request $r$ is not part of multiple assigned request groups in $G$ (Equation \ref{eq:ILP-ass-cstr-2}), and that each decision variable is binary (Equation \ref{eq:ILP-ass-cstr-3}).

Vehicles that are not assigned to a pick-up, can still be assigned to move in the direction of unassigned requests, in anticipation of new requests appearing in areas that currently are under-supplied (step 8 in Fig. \ref{fig:Framework}). In this study, the rebalancing procedure of \citeA{Alonso-Mora2017ondemand} is adopted. Its objective is to minimise the total empty vehicle rebalancing distance while ensuring a maximum number of vehicles to be assigned to rebalance.

After vehicles are assigned to pick-up requests, rebalance or remain idle, vehicle schedules are updated and the simulation prepares for the next assignment phase (step 9 in Fig. \ref{fig:Framework}).

\subsection{Key Performance Indicators}
We measure the performance of the ride-pooling service using a series of Key Performance Indicators (KPIs) designed to capture the level of service (LoS) offered to users as well as its operational efficiency which is relevant for authorities and service providers. If ride-pooling services are assessed by examining the same aspects that public transport users consider to be most important \cite{bates2001valuation, edvardsson1998causes,hensher2003service,konig2002reliability,friman2001frequency}, then the KPIs of shared rides LoS are reliability, comfort, travel time and fare level. Ride fares in this case are not considered as a KPI, since they are directly dependent on $\pi_r$ and are thus endogenously defined as model input. The main LoS KPIs in this study include the acceptance rate (i.e. the percentage of fulfilled requests out of the total demand, thereby an indicator for coverage and reliability), the delay as a percentage of the direct travel time (indicating travel time), the average number of stops per passenger (pertaining to comfort), and the share of passenger time with a specific number of co-riders on-board (also related to comfort).

In addition to the quality of service delivered by the ride-pooling service, an authority is also interested in the share of the vehicle distance that can be reduced through ride-pooling. A suitable KPI to express distance efficiency is the gross effective vehicle transportation distance ratio, which is defined as the sum of the shortest OD-distance of accepted requests (= 'effective vehicle distance') divided by the total vehicle movement distance \cite{bestmile2018simulation}. The total vehicle movement distance (or vehicle mileage) consists of the transportation distance (the vehicle distance with at least one passenger on-board) and the deadheading distance (the total empty vehicle distance for accessing requests and rebalancing). Further, a net effective vehicle transportation distance ratio is defined. This ratio accounts for the fact that the summed shortest distance of accepted requests, which represents the distance needed when sharing is not allowed, excludes deadheading. For a more fair comparison, the deadheading distance is therefore subtracted from the total vehicle movement in the net effective vehicle transportation distance ratio. For operators, the average vehicle occupancy while a vehicle is transporting passengers is an important efficiency KPI.

\section{Experimental design}\label{sect:exp-design}
\noindent A series of experiments is constructed to test the effect of users' behavioural preferences, the discounting policy and the spatial distribution of demand on ride-pooling performance in an urban context. Before specifying the scenarios that have been designed to test the effect of these variables (Subsection \ref{subsect:scenarios}), we introduce the general set-up of the experiment in terms of road network, demand and vehicle fleet characteristics in Subsection \ref{subsect:setup}. Subsection \ref{subsect:implement} concludes this section with a description of the model implementation.

\subsection{Set-up}\label{subsect:setup}
\noindent The assumed grid network consists of 121 nodes with a link distance of 500 meters, thereby leading to a maximum trip distance of 10 kilometres and a surface area of 25 km\textsuperscript{2}, comparable to the area inside the Ring Road of Amsterdam or the Inner Ring of Berlin. The intermediate stop distance is relatively large, whereby we implicitly assume that vehicles cannot stop at all road intersections and users are willing to walk to a pick-up location (and/or from a drop-off location). The assumed speed on the roads is slightly higher than in an average European city \cite{kfzteile2019best}: 36 km/h.

The total demand for trips is set to 1,210 requests per hour, an average of 10 requests per hour per node. The way trips are distributed over the network is scenario-specific, since we are explicitly interested in investigating the impact of demand distribution on system performance. In all cases trips with a ride distance of 2 kilometres or shorter are excluded, as such rides are uncommon \cite{liu2019canpaxflow} as well as undesirable in the context of a ride-pooling service. A gravity model \cite{erlander1990gravity} is applied to create a list of origin-destination pairs. Demand generation is assumed to follow a random process with Poisson distribution. Each request $r$ is assigned with a request time $t_r^r$ by sampling from an exponential distribution based on the expected interval $\lambda$ between two successive requests with a specific OD-combination, which follows again from the (scenario-specific) demand distribution.  

The fleet of the investigated ride-pooling service consists of 150 vehicles with a capacity of $\kappa=3$, that of a normal car, initially evenly distributed over the network. Ride fares are set based on the regulated maximum taxi fares for the city of Amsterdam in 2019: a base fee of \euro{3} and a kilometre fee of \euro{2} \cite{ams2019taxi}. 

For computational reasons, the total duration of the simulation is limited to two hours, with request groups being assigned to vehicles every minute (120 times in total). An additional warm-up period of 15 minutes is applied to minimise the impact of each of the starting conditions.

\subsection{Scenarios}\label{subsect:scenarios}
A total of fourteen scenarios are constructed for investigating the effect of previously introduced behavioural attributes, the platform's pricing policy and the spatial distribution of demand in the network. We limit the number of scenarios in the experiment by designing scenarios that differ only in one variable at a time from a reference scenario. In this sub section, we provide the motivation for the specification of scenarios, summarised in Table \ref{tab:scenarios}, based on the four experimental variables: delay aversion $\beta_r$, reluctance to share $\gamma_r$, sharing discount $\pi_r$, and directionality in demand.

\begin{table}[!ht]
    \footnotesize
    \centering
    \begin{threeparttable}
	\caption{Scenario design ($\alpha_r = 0.5 \cdot \beta_r$).}\label{tab:scenarios}
		\begin{tabular}{l l l l l l}
             \# & Demand distribution & $\beta_r$ (\euro/h) & $\gamma_r$ (\euro) & $\pi_r$ (\%) & Acronym \\\hline
             1 & Uniform & 30 & 3 & 50 & U\_30\_3\_50 \\
             2 & Uniform & 18 & 3 & 50 & U\_18\_3\_50 \\
             3 & Uniform & 24 & 3 & 50 & U\_24\_3\_50 \\
             4 & Uniform & 36 & 3 & 50 & U\_36\_3\_50 \\
             5 & Uniform & 42 & 3 & 50 & U\_42\_3\_50 \\
             6 & Uniform & 30 & 1 & 50 & U\_30\_1\_50 \\
             7 & Uniform & 30 & 2 & 50 & U\_30\_2\_50 \\
             8 & Uniform & 30 & 4 & 50 & U\_30\_4\_50 \\
             9 & Uniform & 30 & 5 & 50 & U\_30\_5\_50 \\
             10 & Uniform & $\mathcal{N}$(30,10) & 3 & 50 & U\_H\_3\_50 \\
             11 & Uniform & 30 & $\mathcal{N}$(3,2) & 50 & U\_30\_H\_50 \\
             12 & Uniform & 30 & 3 & 50 + $7.5 \cdot n^{pax}$  & U\_30\_3\_D \\
             13 & Moderately directed & 30 & 3 & 50 & MD\_30\_3\_50 \\
             14 & Strongly directed & 30 & 3 & 50 & SD\_30\_3\_50 \\\hline
		\end{tabular}
% 	\end{center}
	\end{threeparttable}
\end{table}

\subsubsection{Delay aversion}
Detours for picking up or dropping off additional travellers induce a cost for travellers already on-board the vehicle. Thus, a potentially crucial behavioural factor for the potential to pool rides is the amount of disbenefit that travellers allocate to a single unit of delay, which is represented in this work by delay aversion $\beta_r$ (Equation \ref{eq:benefit}). Several recent studies \cite{krueger2016preferences,lavieri2019modeling,liu2019framework,alonso2020value} have estimated the value of in-vehicle time for ride-pooling, providing empirical indications for the range of travellers' delay aversion values. Their estimates vary considerably, ranging from 8 to 23 \euro{}/h, which implies that the value of in-vehicle time for ride-pooling may be highly context-specific, depending on factors like the type of vehicle (autonomous or humanly-driven) and socio-economic characteristics. Since the specification of delay aversion in ride-pooling is not yet well-established, we decide to test a relatively wide range of values in order to get a more complete picture of how travellers' perception of delays can possibly affect ride-pooling potential. For the base scenario, we assume a delay aversion $\beta_r$ of \euro{30}/h, while alternative scenarios test values that are 6 and 12 \euro{}/h higher or lower than the base value (scenarios 1-5 in Table \ref{tab:scenarios}). 

At the same time, taste variations may play an important role in determining the scalability and efficiency of ride-pooling systems when users are able to choose between private and shared rides, as modelled in this study. \citeA{alonso2020value} found a significant heterogeneity in the value of in-vehicle time of potential ride-pooling users. We therefore introduce a scenario with taste heterogeneity in delay aversion $\beta_r$. Since there is limited empirical knowledge on the variation of delay aversion amongst the population, we need to make an assumption about the shape of this distribution. We specify heterogeneity in delay aversion (scenario 10 in Table \ref{tab:scenarios}) as a normal distribution with a mean value of \euro{30}/h (the base value from scenario 1) and a standard deviation of \euro{10}/h: $\mathcal{N}(30,10)$.

In all of the experiments, the value of $\alpha_r$, representing the extent to which waiting time is perceived more negatively than in-vehicle time, is set to half of $\beta_r$, which is in line with the waiting time multiplier found in stated preference studies on potential ride-pooling users \cite{liu2019framework,alonso2020value} and a revealed preference study in urban public transit \cite{yap2018crowding}.

% Delay aversion might be best compared to the value of travel time reliability (VOR) in public transit. A study on travel time reliability for commuters in Barcelona \cite{asensio2008commuters} found a VOR of \euro{34.4}/h and a similar study for Australia \cite{li2010willingness} concluded a mean value of approximately \euro{33}/h.

% Previous research has found that preference heterogeneity exists regarding adoption of different new mobility services (Alemi et al., 2019, Alonso-González et al., 2020b, El Zarwi et al., 2017), and pooled on-demand services in particular

%   As can be seen, the models suggest that waiting time is valued about one-half as much as in-vehicle travel time. This finding seems to be inconsistent with previous research findings that suggested that travelers place greater value on waiting time than in-vehicle travel (Chavis)

\subsubsection{Reluctance to share}
Recently, the concept of willingness to share, also referred to as the reluctance to share or the willingness to pay to not have to share a ride with other travellers, has started to gain more attention in the literature. \citeA{lavieri2019modeling} concluded that travellers' reluctance to share represents a fixed cost, independent on the duration of the ride. \citeA{alonso2020determinants} confirm this principle, yet with the notation that it applies only to ride-pooling with one or two co-riders, and not for all travellers in the population. When travelling with more than two co-riders, which represents microtransit rather than ride-pooling, they find a travel time dependent willingness to share. For small-scale ride-pooling however, both studies find a relatively small fixed average reluctance to share, in the range of \euro{0.50} - \euro{1}. \citeA{alonso2020determinants} also observe that the willingness to share is highly context-specific, depending on for example geographical characteristics and familiarity with ride-pooling services.

All in all, empirical research on travellers' reluctance to share is still scarce, especially when considering different contexts. With a first indication that the reluctance to share a vehicle with one or two co-riders is represented by a fixed cost, we decide to test a relatively large range of fixed values for reluctance to share $\gamma_r$ in our numerical experiments: from \euro{1} to \euro{5} (scenarios 1 and 6-9 in Table \ref{tab:scenarios}). In all of the other scenarios, we assume a median value of $\gamma_r$ = \euro{3}. Again, when lacking full understanding of the specification of behavioural attributes related to sharing, covering a large range of values allows examining the implications of these attributes for the potential of ride-pooling.

As mentioned earlier, \citeA{alonso2020determinants} state that different classes of potential ride-pooling users have a different specification of their reluctance to share. In fact, preferences related to fellow passengers in ride-pooling are found to exercise more pronounced heterogeneity than preferences towards travel time \cite{zhang2018mobility}. A possible explanation for this is that some users enjoy the social interactions that come along with sharing a ride while others are reluctant to share their ride with strangers. This implies that for some users in fact the wilingness to share $\gamma_r$ might be positive, or in other words that the fact that a vehicle is shared induces an additional benefit next to a reduced ride fare. Again, lacking empirical evidence on the distribution of preferences across the population, a normal distribution is assumed to capture heterogeneity. Since heterogeneity in reluctance to share $\gamma_r$ was found to be (likely) larger than heterogeneity in delay aversion $\beta_r$ (with its assumed standard deviation of one third of the mean), the standard deviation relative to the mean to describe heterogeneity in the reluctance to share $\gamma_r$ is therefore set higher than for the delay aversion $\beta_r$ in scenario 10. We specify a mean reluctance to share of \euro{3} and a standard deviation of \euro{2}: $\mathcal{N}(3,2)$ (scenario 11).

%  This threshold is however surpassed for the four co-rider situation, leading not only to a higher value but to a per-minute value (Alonso-Gonzalez, 2020, Transportation)
%  When pooling rides, the disutility of having one or two extra passengers is constant, regardless of the trip length. This disutility further increases in the event that one shares the ride with four additional passengers, in which case it increases the longer the pooled trip

% These classes have diferent specifcations to represent WTS, indicating that the disutility attributed to sharing is perceived diferently among individuals. 

\subsubsection{Sharing discount}
An additional scenario (scenario 12 in Table \ref{tab:scenarios}) has been devised to test the effect of a ride-pooling providers' pricing mechanism. All scenarios except scenario 12 assume a fixed 50\% discount for all ride-pooling rides, independent of whether sharing actually occurs throughout the ride. This is in line with services offered by ride-pooling platforms, e.g. UberPool, that typically offer discounts between 25 and 60\% of the fare of a private ride \cite{shaheen2019shared}. We are interested in examining the impacts of an alternative pricing mechanism that reflects the actual extent of sharing experienced by the user. We therefore specify an alternative scenario where a similar discount of 50\% is given to the user even if he or she ends up being served privately (same as for all other scenarios, whereby the discount is basically a compensation for the risk of having to share), while an additional 7.5\% discount is given for each co-rider $n^{pax}$ on-board the vehicle during the part of the ride with highest vehicle occupancy. The aim of this additional scenario is to explore whether occupancy-dependent discounts in ride-pooling can be effective in reducing vehicle mileage by facilitating the matching process, as well as at what user cost. It is not meant to give a specification of the (socially) optimal pricing strategy, which, depending on the results of this study, may however be an interesting topic for future research.

\subsubsection{Demand distribution}
The effect of directionality in demand on the performance of ride-pooling is tested using three different scenarios. In the base scenario (1 in Table \ref{tab:scenarios}) demand is perfectly uniform, with equal production and attraction in each of the nodes. In two additional scenarios (scenarios 13 and 14 in Table \ref{tab:scenarios}), we specify a demand distribution that represents an increasingly concentrated demand pattern, with more production in the outer nodes of the network and more attraction in the central nodes, intended to mimic a morning peak pattern.

% While in scenarios 1-9 the overall delay and waiting time aversion are examined, their levels are assumed constant for all users. Taste variations are expected however to play an important role in determining the scalability and efficiency of ride-pooling systems when users are able to choose between private and shared rides as modelled in this study. We therefore introduce taste heterogeneity in delay aversion $\beta_r$ and reluctance to share $\gamma_r$ in scenarios 10 and 11. There is limited empirical knowledge on the variation of  delay aversion and willingness to share values amongst the population. A study on commuters in Shanghai found an average Value of Travel Time Reliability (VOR) of 88.3 Chinese yuan per hour, with the standard deviation measuring 23.9 yuan/h \cite{gao2018heterogeneity}, approximately one third of the average. As the delay aversion compares best to the VOR, we assume a similar ratio for delay aversion $\beta_r$. Heterogeneity in delay aversion (scenario 10 in Table \ref{tab:scenarios}) is thus specified as a normal distribution with a mean value of \euro{30}/h and a standard deviation of \euro{10}/h: $\mathcal{N}(30,10)$. 

\subsection{Implementation}\label{subsect:implement}
The simulation model is implemented in Python from scratch, using the open-source library Numpy to enable efficient operations of large data structures in the model, such as creating and storing the edges of RGV-graphs when many requests and vehicles are considered. The Networkx package is used to compute the shortest path between a pair of locations in the road network, after which the corresponding travel time is stored in a look-up table.

We compute the complete RV- and RGV-graphs without imposing a time budget or limits on the number of edges. An exhaustive search is performed to find the optimal stop sequence with which a vehicle can serve one or more requests (Algorithm \ref{Alg:Route feasibility}). The optimisation problems that are part of the group-vehicle assignment and rebalancing procedure are, for all experiments, solved to optimality using the MOSEK Optimizer API. Consequently, each assignment iteration is guaranteed to yield an optimal result.

With these settings, the majority of scenarios could be run within 30 minutes on a single-core 2.30GHz processor. Two noticeable exceptions are the scenarios with the lowest delay aversion $\beta_r$ and lowest reluctance to share $\gamma_r$ with run times of approximately five hours. In these scenarios, as a result of more favourable preferences towards sharing, larger request groups are potentially feasible, hence increasing the solution space. Consequently, it requires significant computational time to test those as the set of possible stop sequences to satisfy such groups is significantly (i.e. more than exponentially) larger than for small request groups. Also the scenario with an occupancy-dependent additional discount (\textit{U\_30\_3\_D}) enlarges the solution space and consequently the computational complexity of this scenario is also relatively high compared to most other scenarios (i.e. a run time of nearly one hour with the same processor).

Even though we investigate real-time ride-pooling, given the offline evaluation nature of this work we do not focus on the computational efficiency of the algorithm in this study. Notwithstanding, the approach adopted in this study is in principle suited for simulation in real-time, as long as we manage to limit the computational load of each assignment iteration. We can think of several changes to the current implementation to do so, including a timeout or a constraint on the number of edges in the development of RV- and RGV-graphs, a timeout when seeking the optimal stop sequence for a group-vehicle combination, and a timeout in the ILP assignment of vehicles to requests. As these implementation directly touch upon the core of the work by \citeA{Alonso-Mora2017ondemand}, we refer to their work for more details.

% Therefore, instead we mimic real-time ride-pooling operations by halting the clock from the moment that matching commences until it is finalized and vehicles are assigned to requests. By excluding the computational time needed for assignment from the simulation, we significantly reduce the need for a computationally efficient implementation. 

\section{Results}\label{sect:results}
In this section we report the results of the experiment and analyse the effect of users' preferences (Subsection \ref{subsect:hom}), the variation thereof (Subsection \ref{subsect:het}), the applied discount structure (Subsection \ref{subsect:disc}) and the demand distribution (Subsection \ref{subsect:dem}) on the level of service and efficiency of a ride-pooling service. The complete set of KPI values is presented in Tables \ref{tab:KPI-Los-results} (level of service) and \ref{tab:KPI-eff-results} (efficiency). In the following four subsections we discuss the effect of each of the above mentioned aspects in detail. 

\begin{table}[!ht]
\centering
\begin{threeparttable}
\caption{Level of service KPI values for each scenario.}\label{tab:KPI-Los-results}
\scriptsize
\begin{tabular}{@{}lcS[table-format=3.1]S[table-format=3.1]S[table-format=3.1]ccccc@{}}
Scenario & \multicolumn{1}{c}{\rot{\begin{tabular}[c]{@{}l@{}}Acceptance\\ rate \end{tabular}}} & \multicolumn{1}{c}{\rot{\begin{tabular}[c]{@{}l@{}}Average total delay \\ per passenger (s) \end{tabular}}} & \multicolumn{1}{c}{\rot{\begin{tabular}[c]{@{}l@{}}Average waiting time \\ per passenger (s)\end{tabular}}} & \multicolumn{1}{c}{\rot{\begin{tabular}[c]{@{}l@{}}Average in-vehicle delay\\ per passenger (s)\end{tabular}}} & \multicolumn{1}{c}{\rot{\begin{tabular}[c]{@{}l@{}}Average passenger ratio \\ delay / direct travel time\end{tabular}}} & \multicolumn{1}{c}{\rot{\begin{tabular}[c]{@{}l@{}}Average number of stops \\ per passenger \end{tabular}}} & \multicolumn{1}{c}{\rot{\begin{tabular}[c]{@{}l@{}}Ratio of passenger time\\ with 0 co-riders \end{tabular}}} & \multicolumn{1}{c}{\rot{\begin{tabular}[c]{@{}l@{}}Ratio of passenger time \\ with 1 co-rider \end{tabular}}} & \multicolumn{1}{c}{\rot{\begin{tabular}[c]{@{}l@{}}Ratio of passenger time \\ with 2 co-riders \end{tabular}}} \\\hline
U\_30\_3\_50 & 76\% & 126.8 & 92.1 & 34.7 & 30\% & 0.95 & 50\% & 32\% & 18\% \\
U\_18\_3\_50 & 88\% & 200.5 & 122.4 & 78.1 & 49\% & 1.56 & 31\% & 33\% & 36\% \\
U\_24\_3\_50 & 81\% & 158.5 & 103.6 & 55.0 & 37\% & 1.24 & 40\% & 35\% & 25\% \\
U\_36\_3\_50 & 70\% & 106.2 & 84.8 & 21.4 & 25\% & 0.79 & 58\% & 31\% & 11\% \\
U\_42\_3\_50 & 64\% & 93.2 & 77.1 & 16.0 & 21\% & 0.67 & 63\% & 29\% & 8\% \\
U\_30\_1\_50 & 99\% & 219.5 & 138.2 & 81.3 & 61\% & 1.83 & 25\% & 38\% & 37\% \\
U\_30\_2\_50 & 98\% & 169.6 & 114.0 & 55.6 & 45\% & 1.40 & 35\% & 37\% & 28\% \\
U\_30\_4\_50 & 46\% & 102.7 & 82.9 & 19.8 & 21\% & 0.66 & 64\% & 27\% & 9\% \\
U\_30\_5\_50 & 25\% & 88.0 & 74.7 & 13.2 & 15\% & 0.39 & 76\% & 21\% & 3\% \\
U\_H\_3\_50 & 76\% & 154.2 & 103.5 & 50.8 & 36\% & 1.08 & 46\% & 32\% & 22\% \\
U\_30\_H\_50 & 67\% & 162.8 & 106.4 & 56.4 & 42\% & 1.19 & 45\% & 34\% & 21\% \\
U\_30\_3\_D & 82\% & 221.7 & 118.7 & 103.0 & 54\% & 1.97 & 23\% & 32\% & 45\% \\
MD\_30\_3\_50 & 68\% & 117.7 & 92.5 & 25.1 & 29\% & 0.79 & 56\% & 30\% & 14\% \\
SD\_30\_3\_50 & 63\% & 130.7 & 106.5 & 24.2 & 32\% & 0.80 & 53\% & 33\% & 14\% \\ \hline
\end{tabular}
\end{threeparttable}
\end{table}

\begin{table}[!ht]
\centering
\begin{threeparttable}
\caption{Efficiency KPI values for each scenario.}\label{tab:KPI-eff-results}
\scriptsize
\begin{tabular}{@{}lccccccccccc@{}}
%\toprule
Scenario & \multicolumn{1}{c}{\rot{\begin{tabular}[c]{@{}l@{}}Total vehicle \\ movement distance (km) \end{tabular}}} & \multicolumn{1}{c}{\rot{\begin{tabular}[c]{@{}l@{}}Total vehicle \\ transportation distance (km) \end{tabular}}} & \multicolumn{1}{c}{\rot{\begin{tabular}[c]{@{}l@{}}Total \\ deadheading distance (km) \end{tabular}}} & \multicolumn{1}{c}{\rot{\begin{tabular}[c]{@{}l@{}}Total \\ rebalancing distance (km) \end{tabular}}} & \multicolumn{1}{c}{\rot{\begin{tabular}[c]{@{}l@{}}Gross effective vehicle \\ transportation distance ratio \end{tabular}}} & \multicolumn{1}{c}{\rot{\begin{tabular}[c]{@{}l@{}}Net effective vehicle \\ transportation distance ratio \end{tabular}}} & \multicolumn{1}{c}{\rot{\begin{tabular}[c]{@{}l@{}}Empty vehicle rebalancing \\ distance ratio \end{tabular}}} & \multicolumn{1}{c}{\rot{\begin{tabular}[c]{@{}l@{}}Average \\ vehicle occupancy \end{tabular}}} & \multicolumn{1}{c}{\rot{\begin{tabular}[c]{@{}l@{}}Ratio of non-empty \\ vehicle time with occupancy 1 \end{tabular}}} & \multicolumn{1}{c}{\rot{\begin{tabular}[c]{@{}l@{}}Ratio of non-empty \\ vehicle time with occupancy 2 \end{tabular}}} & \multicolumn{1}{c}{\rot{\begin{tabular}[c]{@{}l@{}}Ratio of non-empty \\ vehicle time with occupancy 3 \end{tabular}}} \\\hline
U\_30\_3\_50 & 6,488 & 5,588 & 900 & 156 & 1.15 & 1.34 & 0.173 & 1.38 & 70\% & 22\% & 8\% \\
U\_18\_3\_50 & 6,357 & 5,690 & 667 & 78 & 1.27 & 1.42 & 0.117 & 1.68 & 52\% & 28\% & 20\% \\
U\_24\_3\_50 & 6,456 & 5,657 & 799 & 133 & 1.21 & 1.38 & 0.166 & 1.52 & 61\% & 26\% & 13\% \\
U\_36\_3\_50 & 6,350 & 5,453 & 898 & 156 & 1.12 & 1.30 & 0.174 & 1.30 & 75\% & 20\% & 5\% \\
U\_42\_3\_50 & 6,142 & 5,218 & 925 & 187 & 1.09 & 1.28 & 0.202 & 1.24 & 79\% & 18\% & 3\% \\
U\_30\_1\_50 & 6,389 & 5,867 & 522 & 9 & 1.36 & 1.48 & 0.017 & 1.78 & 44\% & 34\% & 22\% \\
U\_30\_2\_50 & 6,853 & 6,133 & 720 & 56 & 1.26 & 1.41 & 0.077 & 1.59 & 56\% & 30\% & 15\% \\
U\_30\_4\_50 & 4,966 & 4,330 & 636 & 106 & 1.10 & 1.26 & 0.167 & 1.24 & 79\% & 17\% & 4\% \\
U\_30\_5\_50 & 3,321 & 2,958 & 364 & 37 & 1.05 & 1.18 & 0.102 & 1.14 & 87\% & 12\% & 1\% \\
U\_H\_3\_50 & 6,389 & 5,556 & 833 & 152 & 1.16 & 1.33 & 0.182 & 1.45 & 66\% & 23\% & 11\% \\
U\_30\_H\_50 & 5,570 & 4,949 & 621 & 84 & 1.17 & 1.32 & 0.135 & 1.45 & 65\% & 25\% & 10\% \\
U\_30\_3\_D & 5,687 & 5,186 & 501 & 125 & 1.35 & 1.48 & 0.249 & 1.85 & 43\% & 29\% & 28\% \\
MD\_30\_3\_50 & 6,000 & 4,829 & 1,171 & 393 & 1.06 & 1.31 & 0.335 & 1.32 & 74\% & 20\% & 6\% \\
SD\_30\_3\_50 & 6,288 & 4,552 & 1,736 & 910 & 0.95 & 1.31 & 0.524 & 1.35 & 72\% & 22\% & 6\% \\\hline
\end{tabular}
\end{threeparttable}
\end{table}

\subsection{Effect of (homogeneous) behavioural preferences}\label{subsect:hom}
\begin{figure}[!ht]
\captionsetup{justification=centering}
\centering
\begin{minipage}{.4\textwidth}
  \centering
  \includegraphics[width=\linewidth]{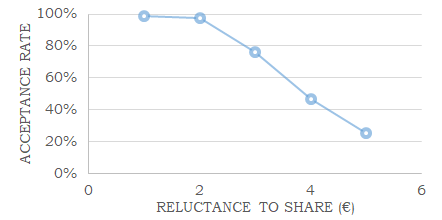}
  \textbf{(a)}
\end{minipage}%
\begin{minipage}{.04\textwidth}
  \includegraphics[width=\linewidth]{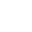}
\end{minipage}%
\begin{minipage}{.4\textwidth}
  \centering
  \includegraphics[width=\linewidth]{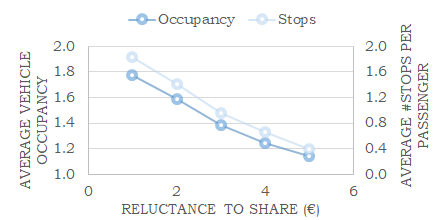}
  \textbf{(b)}
\end{minipage}
\begin{minipage}{.4\textwidth}
  \centering
  \includegraphics[width=\linewidth]{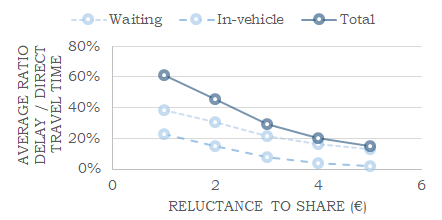}
  \textbf{(c)}
  \label{fig:het_occ-and-stops}
\end{minipage}%
\begin{minipage}{.04\textwidth}
  \includegraphics[width=\linewidth]{Figures/white_space.png}
\end{minipage}%
\begin{minipage}{.4\textwidth}
  \centering
  \includegraphics[width=\linewidth]{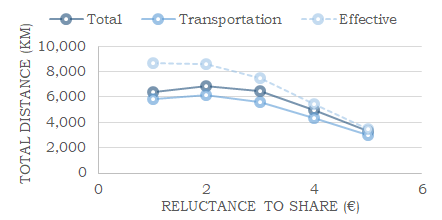}
    \textbf{(d)}
  \label{fig:het_pax-time}
\end{minipage}
\caption{The effect of reluctance to share $\gamma_r$ on (a) acceptance rate, (b) vehicle occupancy and number of intermediate stops, (c) average passenger delay, and (d) total vehicle movement, total transportation distance and effective transportation distance.}
\label{fig:hom_1}
\end{figure}

As can be expected, the acceptance rate (Fig. \ref{fig:hom_1}a) increases as the reluctance to share $\gamma_r$ decreases. The acceptance rate rises from 25.4\% when $\gamma_r$ = \euro{5} to nearly 100\% when $\gamma_r$ = \euro{1}. The increase is approximately linear until the great majority of requests is accepted. Interestingly, the average vehicle occupancy (Fig. \ref{fig:hom_1}b) increases more than linearly when $\gamma_r$ decreases, as well as passengers' waiting time and in-vehicle delay (Fig. \ref{fig:hom_1}c). It is found that hardly any rides are shared (i.e. the average vehicle occupancy is 1.14) if users are highly sensitive to sharing with other passengers, resulting in an average in-vehicle delay close to zero. In such a scenario, the operational efficiency in terms of the number of effective passenger kilometres per vehicle kilometre is as low as 1.05. This ratio is found to increase approximately linearly with an increase in the willingness to share. It can be explained by the finding that the total effective vehicle distance (due to more requests served) increases more than the total vehicle movement distance when users are more flexible (Fig. \ref{fig:hom_1}d), as a result of a more efficient assignment of vehicles to requests. Also, deadheading is found to be relatively uncommon when users' sharing tolerance is high, as new requests can be picked-up by vehicles on their way to drop off other passengers. If $\gamma_r$ = \euro{1} for example, the average effective passenger distance per total vehicle kilometre in the system (including transportation and deadheading) rises to 1.36 kilometre.

When considering the effect of delay aversion $\beta_r$ instead of reluctance to share $\gamma_r$, similar, albeit less pronounced, results are found. The acceptance rate, for example, does not exceed 90\% in any of the scenarios. Evidently, the level of service and operational efficiency are more sensitive to the tested values of the willingness to share, $\gamma_r$, than to those of the delay aversion, $\beta_r$.

\subsection{Effect of taste heterogeneity}\label{subsect:het}
\begin{figure}[!ht]
\centering
\begin{minipage}{.4\textwidth}
  \centering
  \includegraphics[width=\linewidth]{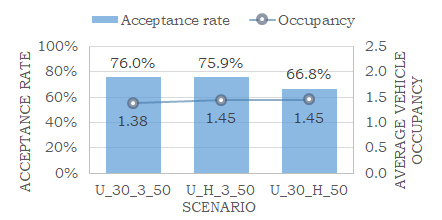}
    \textbf{(a)}
\end{minipage}%
\begin{minipage}{.04\textwidth}
  \includegraphics[width=\linewidth]{Figures/white_space.png}
\end{minipage}%
\begin{minipage}{.4\textwidth}
  \centering
  \includegraphics[width=\linewidth]{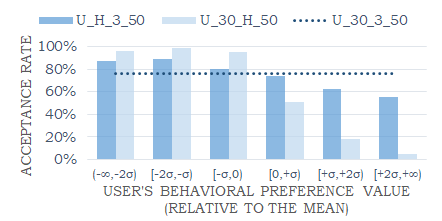}
   \textbf{(b)}
\end{minipage}
\begin{minipage}{.4\textwidth}
  \centering
  \includegraphics[width=\linewidth]{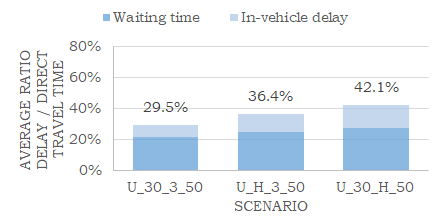}
    \textbf{(c)}
\end{minipage}%
\begin{minipage}{.04\textwidth}
  \includegraphics[width=\linewidth]{Figures/white_space.png}
\end{minipage}%
\begin{minipage}{.4\textwidth}
  \centering
  \includegraphics[width=\linewidth]{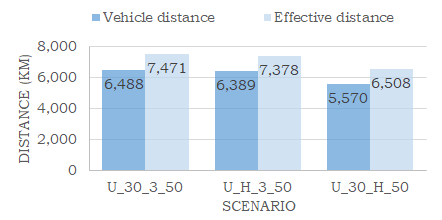}
    \textbf{(d)}
\end{minipage}
\captionsetup{justification=centering}
\caption{The effect of heterogeneity in delay aversion $\beta_r$ and reluctance to share $\gamma_r$ on (a) average vehicle occupancy, (a) and (b) acceptance rate, (c) average passenger delay, and (d) total vehicle movement distance and effective transportation distance (the difference between those indicating distance savings).}
\label{fig:het_1}
\end{figure}

Significantly fewer requests are accepted when reluctance to share varies amongst the user population (with an unchanged mean): 66.8\% versus 76.0\% of all requests (Fig. \ref{fig:het_1}a). As can be seen in Fig. \ref{fig:het_1}b, the acceptance rate varies considerably amongst user groups that are characterised by different degrees of reluctance to share. Heterogeneity in the delay tolerance on the other hand barely has an impact on the acceptance rate: 75.9\%. An explanation for this difference could be that users with a high delay aversion can often still be satisfied (Fig. \ref{fig:het_1}b) by serving them with the shortest or a relatively short path so that their delay costs are minimised. This happens at the cost of more flexible passengers that will get served with a delayed pick-up and/or a less direct route. In other words, a level of service is offered that discriminates amongst users based on their delay aversion. With the majority of accepted requests having an above average delay tolerance (since it is easier to find a ride for those requests), the average delay experienced by users in the system, 36.4\% of the direct travel time, is higher than when the delay tolerance is assumed homogeneous (29.5\%, Fig. \ref{fig:het_1}c). 

When considering a scenario with varying willingness to share, it is found that users with a below average willingness to share are much more likely to reject a ride-pooling service (Fig. \ref{fig:het_1}b) even when offered a direct ride, leading to a lower overall acceptance rate (Fig. \ref{fig:het_1}a). The total rebalancing distance is limited since there are relatively many requests around unassigned vehicles that cannot be served even with a direct route. The effective transportation distance ratio, an indicator for distance savings (Fig. \ref{fig:het_1}d), ranges between 1.15 and 1.17 for the three scenarios in this set of experiments, and thus does not seem to be significantly dependent on whether heterogeneity in ride-pooling tolerances is considered. Based on this, we can say that the operational efficiency in terms of the vehicle-km travelled that a ride-pooling system can save is not very sensitive to the variation of sharing preferences over the population.

\subsection{Effect of discount mechanism}\label{subsect:disc}
\begin{figure}[!ht]
\centering
\begin{minipage}{.4\textwidth}
  \centering
  \includegraphics[width=\linewidth]{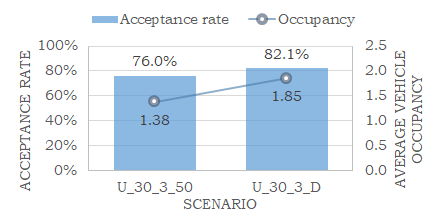}
  \textbf{(a)}
\end{minipage}%
\begin{minipage}{.295\textwidth}
  \centering
  \includegraphics[width=\linewidth]{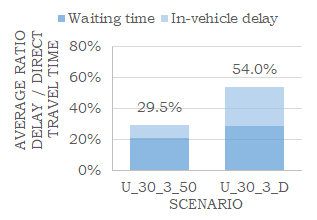}
  \textbf{(b)}
\end{minipage}
\begin{minipage}{.295\textwidth}
  \centering
  \includegraphics[width=\linewidth]{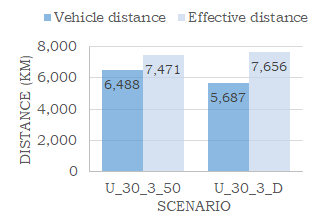}
  \textbf{(c)}
\end{minipage}
\captionsetup{justification=centering}
\caption{The effect of discount structure on (a) acceptance rate and average vehicle occupancy, (b) average passenger delay, and (c) total vehicle movement distance and effective transportation distance (the difference between those indicating distance savings).}
\label{fig:disc_1}
\end{figure}

As expected, when users receive an additional 7.5\% discount per co-rider they share the highest occupancy part of their ride with, the average vehicle occupancy increases dramatically (from 1.38 to 1.85, as can be seen in Fig. \ref{fig:disc_1}a), and a similar increase is found in the share of time passengers spend in a full vehicle (from 17,7\% to 45.1\% of the total passenger time). By utilising the available vehicle capacity more efficiently, the acceptance rate (also Fig. \ref{fig:disc_1}a) increases from 76.0\% to 82.1\%, although at the cost of a higher average delay (Fig. \ref{fig:disc_1}b). A higher vehicle occupancy will burden passengers with longer detours and consequently an in-vehicle delay of more than three times as high as when no additional discount is offered (25.1\% vs 8.1\% of the direct travel time). Also, the average waiting time for pick-up is marginally longer in the scenario with an occupancy-dependent discount, with pick-ups being complicated by the fact that many vehicles are driving around fully occupied.

The (gross) effective vehicle transportation distance ratio increases from 1.15 to 1.35 when an additional 7.5\% discount is awarded per co-rider. In combination with a higher acceptance rate, relatively large distance savings (Fig. \ref{fig:disc_1}c) are achieved with the introduction of an additional occupancy-dependent discount. Passengers however may be reluctant to choose a service for which they know the maximum cost but not the exact cost a-priori. The distance that a ride-pooling service can save (when compared to private rides) stems not only from the more efficient transportation of requests (the transportation distance drops from 5,588 to 4,829 kilometres) but also from a reduction in the deadheading distance to access new requests (from 900 to 501 kilometres), as requests are being picked-up by non-empty vehicles on their way to drop off other passengers.

\subsection{Effect of demand distribution}\label{subsect:dem}
\begin{figure}[!ht]
\centering
\begin{minipage}{.4\textwidth}
  \centering
  \includegraphics[width=\linewidth]{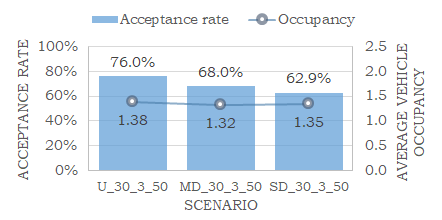}
    \textbf{(a)}
\end{minipage}%
\begin{minipage}{.04\textwidth}
  \includegraphics[width=\linewidth]{Figures/white_space.png}
\end{minipage}%
\begin{minipage}{.4\textwidth}
  \centering
  \includegraphics[width=\linewidth]{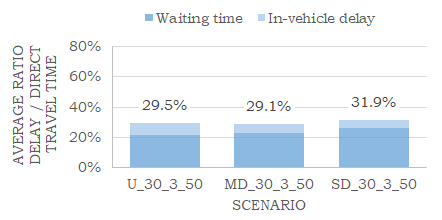}
   \textbf{(b)}
\end{minipage}
\begin{minipage}{.4\textwidth}
  \centering
  \includegraphics[width=\linewidth]{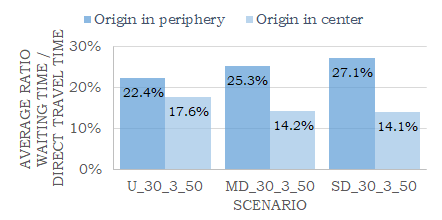}
    \textbf{(c)}
\end{minipage}%
\begin{minipage}{.04\textwidth}
  \includegraphics[width=\linewidth]{Figures/white_space.png}
\end{minipage}%
\begin{minipage}{.4\textwidth}
  \centering
  \includegraphics[width=\linewidth]{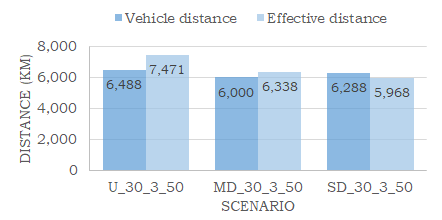}
    \textbf{(d)}
\end{minipage}
\captionsetup{justification=centering}
\caption{The effect of directionality in demand on (a) acceptance rate and average vehicle occupancy, (b) average passenger delay, (c) location-based average waiting time, and (d) total vehicle movement distance and effective transportation distance (the difference between those indicating distance savings).}
\label{fig:dem_1}
\end{figure}

More directionality in demand leads to more requests being rejected by the ride-pooling service, namely 37.1\% when demand is strongly directed versus 24.0\% when demand is perfectly uniform, as shown by Fig. \ref{fig:dem_1}a. If demand is perfectly uniform, the average vehicle occupancy of vehicles in revenue mode (also Fig. \ref{fig:dem_1}a) is 1.38 and the average passenger delay (Fig. \ref{fig:dem_1}b) is 29.5\% of the direct travel time. The drop in the number of accepted requests when there is a moderate level of direction in demand leads to a drop in the vehicle occupancy (1.32) and average delay (29.1\% of direct travel time). Interestingly, when the level of direction increases further however, the vehicle occupancy (1.35) and the average delay (31.9\% of the direct travel time) bounce back. With a larger spatial inequality in pick-ups and drop-offs, average waiting times are relatively short in the center, where attraction exceeds production (Fig. \ref{fig:dem_1}c), compared to the nodes in the periphery of the network. Since only the minority of requests originates in the center in this case, the average waiting time is mainly determined by requests originating outside of the center, where production exceeds attraction. In these nodes, the average waiting time is nearly twice as high (27.1\% versus 14.1\%). 

% \newpage
Deadheading serves as a mean to solve inequality in supply and demand. It is responsible for 27.6\% of all vehicle kilometres in a scenario in which demand is strongly directed, compared to only 13.9\% of the mileage when demand is uniform. The effective passenger kilometres per ride-pooling vehicle kilometre in respective scenarios is 0.95 versus 1.15. This alarming result suggests that when directionality in demand is high, the total vehicle distance can be longer than the effective transportation distance even for ride-pooling services (Fig. \ref{fig:dem_1}d).

\section{Discussion and conclusions}\label{sect:discussion}
This work is the first study to consider ride-pooling potential while accounting for the trade-off that users encounter when presented with the option of ride-pooling. Previous studies, such as \citeA{Santi2014quantifying} and \citeA{Alonso-Mora2017ondemand}, assumed that all users are potentially willing to ride-share as long as their waiting time and total delay do not exceed a certain non-compensatory threshold. This is not very realistic as taxi users have no reason to share their ride (and accept a delay) unless attaining a benefit in return. Therefore, in this study we formulate a compensatory user cost formulation where the disbenefits associated with sharing one's ride need to be at least compensated by the fare reduction offered for the user to substitute a private ride with ride-pooling. Hence, the assumption is that users will only switch to a ride-pooling service if such a choice yields a net positive benefit over a conventional taxi or ride-hailing service. Also, this work accounts for the fact that sharing a vehicle with strangers, which in literature is considered to be one of the main potential barriers for a successful implementation of ride-pooling services, induces a disutility.

Our compensatory formulation is embedded in the the matching framework proposed in \citeA{Alonso-Mora2017ondemand}. A group of requests is considered feasible only if all the individuals included in the shared ride evaluate it as superior to the private ride alternative, as well as satisfying vehicle-related constraints. 

When representing the choice whether to ride-share or not as a compensatory function between travel attributes (travel time, ride fare and the presence of co-riders), we find that the distance savings from ride-pooling reported by previous studies are only attainable when users have favourable preferences to sharing rides. For example, when users have a relatively high willingness to share, or more specifically when they are willing to pay no more than 1 euro to upgrade a shared ride to an individual one, assuming no change in travel time, 32\% of the transportation vehicle kilometres in the network can be removed. This compares to a 40\% reduction found by \citeA{Santi2014quantifying} in their study on ride-pooling in New York. However, our work shows that if users are found to be less willing to share - or in other words, users are willing to pay high premium for a private ride - we can expect distance savings that are significantly lower than reported in previous work. We find for example that, in the scenario where travellers are least willing to share, the total vehicle transportation distance can be reduced by just 15\% when allowing for shared rides. This scenario assumes that ride-pooling users demand a compensation of 5 euro for the fact that they have to share their vehicle with other passengers. Furthermore, it should be noted that in this scenario, and in most others, ride-pooling users get rewarded with a discount rate of 50\%, relatively high compared to discounts currently offered by ride-pooling services. It is thus plausible that ride-pooling distance savings in reality are even lower than those found in our analysis.

In addition, our results show that besides the expected distance savings also the level of service that a ride-pooling system offers is significantly dependent on users' tolerance to delays and willingness to share a vehicle with co-riders. When we vary both attributes, we find the acceptance rate of a ride-pooling service to range between 25\% and 99\% of all incoming requests, and the average delay of accepted requests to range between 15\% and 61\% of the direct travel time. In conclusion, our results highlight that any comparison of ride-pooling model outputs has to carefully account for the behavioural specifications. 

Although heterogeneity in user preferences seems to have only a limited impact on potential distance savings, it is found to have negative consequences for the level of service with more rejected requests and a larger average delay. Moreover, this results in striving to serve few users with a below average sharing tolerance at the expense that many of the accepted requests with an above average tolerance get assigned a less efficient ride. This calls for the development of discriminative pricing or service provision mechanisms, including the possibility for the service provider to choose to decline travel requests.

Furthermore, this study has shown that the design of a ride-pooling service, such as its pricing structure, may significantly affect the expected societal benefits and service quality. A relatively small additional discount of 7.5\% per co-rider with whom a user shares their ride at maximum occupancy, on top of the standard 50\% discount assumed in all experiments, can for example more than double the total reduction in vehicle kilometres. At the same time, the percentage of rejected requests drops from 24\% to 18\% if such a discount policy is implemented. Instead of offering a flat subsidy per pooled ride or charging a tax on private rides, policy makers looking to minimise road traffic externalities may thus be better off subsidising ride-pooling based on vehicle occupancy.

% Hence, the pricing structure of alternative scenarios can have substantial consequences for both service performance and related externalities.
% In return for a discount, users are on average burdened with an extra travel time of 24.5\% of the direct travel time. 

This study also shows that the potential of a ride-pooling system can be greatly dependent on external variables, such as the spatial distribution of demand. Demand patterns are likely to be at least somewhat concentrated due to spatial clustering of activities like work, residence and shopping. This study shows for example that, when most requests are directed towards the center of the network, typical for a morning peak, the performance of a ride-pooling system is worse than when demand is uniform, both in terms of level of service and efficiency. This is a direct result of the spatial imbalance in demand and supply, which requires vehicles to deadhead from drop-off locations in low demand areas to pick-up locations in high demand areas, leading to increased vehicle mileage and longer waiting times for pick-up. To be specific, the (gross) effective vehicle transportation distance ratio can even drop below 1 when the directionality of demand is high, while the average user delay can amount to 32\% of the direct travel time in such a scenario. In summary, we found that directionality in demand negatively affects both level of service and operational efficiency of ride-pooling services.

Future research can investigate the impact of additional aspects on ride-pooling performance. This includes for example the effect of fleet properties (capacity and fleet size), the effect of the fares of alternative (single-rider) services, test a more complex discounting mechanism than the one assumed here, or investigate external variables like the density of possible pick-up and drop-off locations in the network. Also, it might be interesting to find whether ride-pooling efficiency can be improved by rejecting requests that negatively affect ride-pooling performance on a system level, such as requests that are destined for a location far away from where new demand is expected. Moreover, the validity of ride-pooling studies can be improved by the incorporation of mode choice, whereby passengers can choose to chose not only between a private ride and ride-pooling, but also choose to travel by other means, possibly adopting a probabilistic choice modelling framework. Finally, future research can address the equity aspects of a ride-pooling system, including the spatial disparity of service accessibility.

\subsection*{Acknowledgements}
\noindent This work was supported by the CriticalMaaS project [grant number 804469], which is financed by the European Research Council and the Amsterdam Institute for Advanced Metropolitan Solutions. A preliminary version of this paper was presented at the 2020 Transportation Research Board Annual Meeting in Washington D.C.

% \noindent This research was supported by the CriticalMaaS project (grant number 804469), which is financed by the European Research Council and the Amsterdam Institute for Advanced Metropolitan Solutions. A preliminary version of this paper was presented at the 2020 Transportation Research Board Annual Meeting in Washington D.C.

\subsection*{Conflicts of interest}
\noindent The authors declare that there are no conflicts of interest in relation to this work.

\subsection*{Contribution}
\noindent The authors confirm contribution to the paper as follows: study conception and design; de Ruijter, Cats; analysis and interpretation of results: de Ruijter, Cats; draft manuscript preparation: de Ruijter, Cats. All authors reviewed the results and approved the final version of the manuscript.

% \newpage

% \nolinenumbers
\bibliographystyle{apacite}
\bibliography{bibliography}

\begin{thebibliography}{}

\bibitem [\protect \citeauthoryear {%
Alonso-Gonz{\'a}lez%
, Cats%
, van Oort%
, Hoogendoorn-Lanser%
\BCBL {}\ \BBA {} Hoogendoorn%
}{%
Alonso-Gonz{\'a}lez%
, Cats%
\BCBL {}\ \protect \BOthers {.}}{%
{\protect \APACyear {2020}}%
}]{%
alonso2020determinants}
\APACinsertmetastar {%
alonso2020determinants}%
\begin{APACrefauthors}%
Alonso-Gonz{\'a}lez, M\BPBI J.%
, Cats, O.%
, van Oort, N.%
, Hoogendoorn-Lanser, S.%
\BCBL {}\ \BBA {} Hoogendoorn, S.%
\end{APACrefauthors}%
\unskip\
\newblock
\APACrefYearMonthDay{2020}{}{}.
\newblock
{\BBOQ}\APACrefatitle {What are the determinants of the willingness to share
  rides in pooled on-demand services?} {What are the determinants of the
  willingness to share rides in pooled on-demand services?}{\BBCQ}
\newblock
\APACjournalVolNumPages{Transportation}{}{}{1--33}.
\PrintBackRefs{\CurrentBib}

\bibitem [\protect \citeauthoryear {%
Alonso-Gonz{\'a}lez%
, van Oort%
, Cats%
, Hoogendoorn-Lanser%
\BCBL {}\ \BBA {} Hoogendoorn%
}{%
Alonso-Gonz{\'a}lez%
, van Oort%
\BCBL {}\ \protect \BOthers {.}}{%
{\protect \APACyear {2020}}%
}]{%
alonso2020value}
\APACinsertmetastar {%
alonso2020value}%
\begin{APACrefauthors}%
Alonso-Gonz{\'a}lez, M\BPBI J.%
, van Oort, N.%
, Cats, O.%
, Hoogendoorn-Lanser, S.%
\BCBL {}\ \BBA {} Hoogendoorn, S.%
\end{APACrefauthors}%
\unskip\
\newblock
\APACrefYearMonthDay{2020}{}{}.
\newblock
{\BBOQ}\APACrefatitle {Value of time and reliability for urban pooled on-demand
  services} {Value of time and reliability for urban pooled on-demand
  services}.{\BBCQ}
\newblock
\APACjournalVolNumPages{Transportation Research Part C: Emerging
  Technologies}{115}{}{102621}.
\PrintBackRefs{\CurrentBib}

\bibitem [\protect \citeauthoryear {%
Alonso-Mora%
, Samaranayake%
, Wallar%
, Frazzoli%
\BCBL {}\ \BBA {} Rus%
}{%
Alonso-Mora%
\ \protect \BOthers {.}}{%
{\protect \APACyear {2017}}%
}]{%
Alonso-Mora2017ondemand}
\APACinsertmetastar {%
Alonso-Mora2017ondemand}%
\begin{APACrefauthors}%
Alonso-Mora, J.%
, Samaranayake, S.%
, Wallar, A.%
, Frazzoli, E.%
\BCBL {}\ \BBA {} Rus, D.%
\end{APACrefauthors}%
\unskip\
\newblock
\APACrefYearMonthDay{2017}{}{}.
\newblock
{\BBOQ}\APACrefatitle {On-demand high-capacity ride-sharing via dynamic
  trip-vehicle assignment} {On-demand high-capacity ride-sharing via dynamic
  trip-vehicle assignment}.{\BBCQ}
\newblock
\APACjournalVolNumPages{Proceedings of the National Academy of Sciences of the
  United States of America}{114: 462--467}{}{}.
\newblock
\begin{APACrefDOI} \doi{10.1073/pnas.1611675114} \end{APACrefDOI}
\PrintBackRefs{\CurrentBib}

\bibitem [\protect \citeauthoryear {%
Bates%
, Polak%
, Jones%
\BCBL {}\ \BBA {} Cook%
}{%
Bates%
\ \protect \BOthers {.}}{%
{\protect \APACyear {2001}}%
}]{%
bates2001valuation}
\APACinsertmetastar {%
bates2001valuation}%
\begin{APACrefauthors}%
Bates, J.%
, Polak, J.%
, Jones, P.%
\BCBL {}\ \BBA {} Cook, A.%
\end{APACrefauthors}%
\unskip\
\newblock
\APACrefYearMonthDay{2001}{}{}.
\newblock
{\BBOQ}\APACrefatitle {The valuation of reliability for personal travel} {The
  valuation of reliability for personal travel}.{\BBCQ}
\newblock
\APACjournalVolNumPages{Transportation Research Part E: Logistics and
  Transportation Review}{37: 191--229}{}{}.
\PrintBackRefs{\CurrentBib}

\bibitem [\protect \citeauthoryear {%
Cici%
, Markopoulou%
, Fr{\'\i}as-Mart{\'\i}nez%
\BCBL {}\ \BBA {} Laoutaris%
}{%
Cici%
\ \protect \BOthers {.}}{%
{\protect \APACyear {2013}}%
}]{%
cici2013quantifying}
\APACinsertmetastar {%
cici2013quantifying}%
\begin{APACrefauthors}%
Cici, B.%
, Markopoulou, A.%
, Fr{\'\i}as-Mart{\'\i}nez, E.%
\BCBL {}\ \BBA {} Laoutaris, N.%
\end{APACrefauthors}%
\unskip\
\newblock
\APACrefYearMonthDay{2013}{}{}.
\newblock
{\BBOQ}\APACrefatitle {Quantifying the {P}otential of {R}ide-{S}haring {U}sing
  {C}all {D}escription {R}ecords} {Quantifying the {P}otential of
  {R}ide-{S}haring {U}sing {C}all {D}escription {R}ecords}.{\BBCQ}
\newblock
\BIn{} \APACrefbtitle {Proceedings of the 14th {W}orkshop on {M}obile
  {C}omputing {S}ystems and {A}pplications.} {Proceedings of the 14th
  {W}orkshop on {M}obile {C}omputing {S}ystems and {A}pplications.}
\PrintBackRefs{\CurrentBib}

\bibitem [\protect \citeauthoryear {%
{City of Toronto}%
}{%
{City of Toronto}%
}{%
{\protect \APACyear {2019}}%
}]{%
toronto2019transportation}
\APACinsertmetastar {%
toronto2019transportation}%
\begin{APACrefauthors}%
{City of Toronto}.%
\end{APACrefauthors}%
\unskip\
\newblock
\APACrefYearMonthDay{2019}{}{}.
\newblock
\APACrefbtitle {The {T}ransportation {I}mpacts of {V}ehicle-for-{H}ire in the
  {C}ity of {T}oronto} {The {T}ransportation {I}mpacts of {V}ehicle-for-{H}ire
  in the {C}ity of {T}oronto}\ \APACbVolEdTR{}{\BTR{}}.
\PrintBackRefs{\CurrentBib}

\bibitem [\protect \citeauthoryear {%
Correia%
\ \BBA {} Viegas%
}{%
Correia%
\ \BBA {} Viegas%
}{%
{\protect \APACyear {2011}}%
}]{%
Correia2011carpooling}
\APACinsertmetastar {%
Correia2011carpooling}%
\begin{APACrefauthors}%
Correia, G.%
\BCBT {}\ \BBA {} Viegas, J.%
\end{APACrefauthors}%
\unskip\
\newblock
\APACrefYearMonthDay{2011}{}{}.
\newblock
{\BBOQ}\APACrefatitle {Carpooling and carpool clubs: {C}larifying concepts and
  assessing value enhancement possibilities through a {S}tated {P}reference web
  survey in {L}isbon, {P}ortugal} {Carpooling and carpool clubs: {C}larifying
  concepts and assessing value enhancement possibilities through a {S}tated
  {P}reference web survey in {L}isbon, {P}ortugal}.{\BBCQ}
\newblock
\APACjournalVolNumPages{Transportation Research Part A: Policy and
  Practice}{45: 81--90}{}{}.
\newblock
\begin{APACrefDOI} \doi{10.1016/j.tra.2010.11.001} \end{APACrefDOI}
\PrintBackRefs{\CurrentBib}

\bibitem [\protect \citeauthoryear {%
Dueker%
, Bair%
\BCBL {}\ \BBA {} Levin%
}{%
Dueker%
\ \protect \BOthers {.}}{%
{\protect \APACyear {1977}}%
}]{%
Dueker1977ridesharing}
\APACinsertmetastar {%
Dueker1977ridesharing}%
\begin{APACrefauthors}%
Dueker, K\BPBI J.%
, Bair, B\BPBI O.%
\BCBL {}\ \BBA {} Levin, I\BPBI P.%
\end{APACrefauthors}%
\unskip\
\newblock
\APACrefYearMonthDay{1977}{}{}.
\newblock
{\BBOQ}\APACrefatitle {Ride sharing: Psychological factors} {Ride sharing:
  Psychological factors}.{\BBCQ}
\newblock
\APACjournalVolNumPages{Transportation Engineering Journal of the American
  Society of Civil Engineers}{103: 685--692}{}{}.
\PrintBackRefs{\CurrentBib}

\bibitem [\protect \citeauthoryear {%
Edvardsson%
}{%
Edvardsson%
}{%
{\protect \APACyear {1998}}%
}]{%
edvardsson1998causes}
\APACinsertmetastar {%
edvardsson1998causes}%
\begin{APACrefauthors}%
Edvardsson, B.%
\end{APACrefauthors}%
\unskip\
\newblock
\APACrefYearMonthDay{1998}{}{}.
\newblock
{\BBOQ}\APACrefatitle {Causes of customer dissatisfaction-studies of public
  transport by the critical-incident method} {Causes of customer
  dissatisfaction-studies of public transport by the critical-incident
  method}.{\BBCQ}
\newblock
\APACjournalVolNumPages{Managing Service Quality: An International
  Journal}{}{}{8: 189--197}.
\PrintBackRefs{\CurrentBib}

\bibitem [\protect \citeauthoryear {%
Ehsani%
, Guglielmetti%
, Laumanns%
, Markov%
\BCBL {}\ \BBA {} de Souza%
}{%
Ehsani%
\ \protect \BOthers {.}}{%
{\protect \APACyear {2018}}%
}]{%
bestmile2018simulation}
\APACinsertmetastar {%
bestmile2018simulation}%
\begin{APACrefauthors}%
Ehsani, S.%
, Guglielmetti, R.%
, Laumanns, M.%
, Markov, I.%
\BCBL {}\ \BBA {} de Souza, R.%
\end{APACrefauthors}%
\unskip\
\newblock
\APACrefYearMonthDay{2018}{}{}.
\newblock
\APACrefbtitle {Simulation-based design and analysis of on-demand mobility
  services} {Simulation-based design and analysis of on-demand mobility
  services}\ \APACbVolEdTR{}{\BTR{}}.
\newblock
\APACaddressInstitution{}{Bestmile}.
\PrintBackRefs{\CurrentBib}

\bibitem [\protect \citeauthoryear {%
Erlander%
\ \BBA {} Stewart%
}{%
Erlander%
\ \BBA {} Stewart%
}{%
{\protect \APACyear {1990}}%
}]{%
erlander1990gravity}
\APACinsertmetastar {%
erlander1990gravity}%
\begin{APACrefauthors}%
Erlander, S.%
\BCBT {}\ \BBA {} Stewart, N\BPBI F.%
\end{APACrefauthors}%
\unskip\
\newblock
\APACrefYear{1990}.
\newblock
\APACrefbtitle {The {G}ravity {M}odel in {T}ransportation {A}nalysis: {T}heory
  and {E}xtensions} {The {G}ravity {M}odel in {T}ransportation {A}nalysis:
  {T}heory and {E}xtensions}.
\newblock
\APACaddressPublisher{Utrecht}{VSP}.
\PrintBackRefs{\CurrentBib}

\bibitem [\protect \citeauthoryear {%
Fiedler%
, {\v{C}}ertick{\`y}%
, Alonso-Mora%
\BCBL {}\ \BBA {} {\v{C}}{\'a}p%
}{%
Fiedler%
\ \protect \BOthers {.}}{%
{\protect \APACyear {2018}}%
}]{%
fiedler2018impact}
\APACinsertmetastar {%
fiedler2018impact}%
\begin{APACrefauthors}%
Fiedler, D.%
, {\v{C}}ertick{\`y}, M.%
, Alonso-Mora, J.%
\BCBL {}\ \BBA {} {\v{C}}{\'a}p, M.%
\end{APACrefauthors}%
\unskip\
\newblock
\APACrefYearMonthDay{2018}{}{}.
\newblock
{\BBOQ}\APACrefatitle {The {I}mpact of {R}idesharing in {M}obility-on-{D}emand
  {S}ystems: {S}imulation {C}ase {S}tudy in {P}rague} {The {I}mpact of
  {R}idesharing in {M}obility-on-{D}emand {S}ystems: {S}imulation {C}ase
  {S}tudy in {P}rague}.{\BBCQ}
\newblock
\BIn{} \APACrefbtitle {{IEEE International Conference on Intelligent
  Transportation Systems}.} {{IEEE International Conference on Intelligent
  Transportation Systems}.}
\PrintBackRefs{\CurrentBib}

\bibitem [\protect \citeauthoryear {%
Friman%
\ \BBA {} G{\"a}rling%
}{%
Friman%
\ \BBA {} G{\"a}rling%
}{%
{\protect \APACyear {2001}}%
}]{%
friman2001frequency}
\APACinsertmetastar {%
friman2001frequency}%
\begin{APACrefauthors}%
Friman, M.%
\BCBT {}\ \BBA {} G{\"a}rling, T.%
\end{APACrefauthors}%
\unskip\
\newblock
\APACrefYearMonthDay{2001}{}{}.
\newblock
{\BBOQ}\APACrefatitle {Frequency of negative critical incidents and
  satisfaction with public transport services. {II}} {Frequency of negative
  critical incidents and satisfaction with public transport services.
  {II}}.{\BBCQ}
\newblock
\APACjournalVolNumPages{Journal of Retailing and Consumer Services}{8:
  105--114}{}{}.
\PrintBackRefs{\CurrentBib}

\bibitem [\protect \citeauthoryear {%
{Gemeente Amsterdam}%
}{%
{Gemeente Amsterdam}%
}{%
{\protect \APACyear {2019}}%
}]{%
ams2019taxi}
\APACinsertmetastar {%
ams2019taxi}%
\begin{APACrefauthors}%
{Gemeente Amsterdam}.%
\end{APACrefauthors}%
\unskip\
\newblock
\APACrefYearMonthDay{2019}{}{}.
\newblock
\APACrefbtitle {Wat kost een rit in een taxi?} {Wat kost een rit in een taxi?}
\newblock
\begin{APACrefURL} [{January 23,
  2019}]\url{https://www.amsterdam.nl/veelgevraagd/?caseid=\%7B7DC3ADE8-163A-4201-ABC4-4ACC1830EB88\%7D}
  \end{APACrefURL}
\PrintBackRefs{\CurrentBib}

\bibitem [\protect \citeauthoryear {%
Hensher%
}{%
Hensher%
}{%
{\protect \APACyear {2020}}%
}]{%
covid19maas}
\APACinsertmetastar {%
covid19maas}%
\begin{APACrefauthors}%
Hensher, D\BPBI A.%
\end{APACrefauthors}%
\unskip\
\newblock
\APACrefYearMonthDay{2020}{}{}.
\newblock
{\BBOQ}\APACrefatitle {What might Covid-19 mean for mobility as a service
  (MaaS)?} {What might covid-19 mean for mobility as a service (maas)?}{\BBCQ}
\newblock
\APACjournalVolNumPages{Transport Reviews}{40}{5}{551-556}.
\newblock
\begin{APACrefURL} \url{https://doi.org/10.1080/01441647.2020.1770487}
  \end{APACrefURL}
\newblock
\begin{APACrefDOI} \doi{10.1080/01441647.2020.1770487} \end{APACrefDOI}
\PrintBackRefs{\CurrentBib}

\bibitem [\protect \citeauthoryear {%
Hensher%
, Stopher%
\BCBL {}\ \BBA {} Bullock%
}{%
Hensher%
\ \protect \BOthers {.}}{%
{\protect \APACyear {2003}}%
}]{%
hensher2003service}
\APACinsertmetastar {%
hensher2003service}%
\begin{APACrefauthors}%
Hensher, D\BPBI A.%
, Stopher, P.%
\BCBL {}\ \BBA {} Bullock, P.%
\end{APACrefauthors}%
\unskip\
\newblock
\APACrefYearMonthDay{2003}{}{}.
\newblock
{\BBOQ}\APACrefatitle {Service quality - developing a service quality index in
  the provision of commercial bus contracts} {Service quality - developing a
  service quality index in the provision of commercial bus contracts}.{\BBCQ}
\newblock
\APACjournalVolNumPages{Transportation Research Part A: Policy and
  Practice}{37: 499--517}{}{}.
\PrintBackRefs{\CurrentBib}

\bibitem [\protect \citeauthoryear {%
Kfzteile24%
}{%
Kfzteile24%
}{%
{\protect \APACyear {2019}}%
}]{%
kfzteile2019best}
\APACinsertmetastar {%
kfzteile2019best}%
\begin{APACrefauthors}%
Kfzteile24.%
\end{APACrefauthors}%
\unskip\
\newblock
\APACrefYearMonthDay{2019}{}{}.
\newblock
\APACrefbtitle {Best and Worst Cities to Drive 2017.} {Best and worst cities to
  drive 2017.}
\newblock
\begin{APACrefURL} [{January 23,
  2019}]\url{https://www.kfzteile24.de/best-and-worst-cities-to-drive-usd}
  \end{APACrefURL}
\PrintBackRefs{\CurrentBib}

\bibitem [\protect \citeauthoryear {%
K{\"o}nig%
\ \BBA {} Axhausen%
}{%
K{\"o}nig%
\ \BBA {} Axhausen%
}{%
{\protect \APACyear {2002}}%
}]{%
konig2002reliability}
\APACinsertmetastar {%
konig2002reliability}%
\begin{APACrefauthors}%
K{\"o}nig, A.%
\BCBT {}\ \BBA {} Axhausen, K\BPBI W.%
\end{APACrefauthors}%
\unskip\
\newblock
\APACrefYearMonthDay{2002}{}{}.
\newblock
{\BBOQ}\APACrefatitle {The {R}eliability of the {T}ransportation {S}ystem and
  its {I}nfluence on the {C}hoice {B}ehaviour} {The {R}eliability of the
  {T}ransportation {S}ystem and its {I}nfluence on the {C}hoice
  {B}ehaviour}.{\BBCQ}
\newblock
\BIn{} \APACrefbtitle {2nd {S}wiss {T}ransport {R}esearch {C}onference.} {2nd
  {S}wiss {T}ransport {R}esearch {C}onference.}
\PrintBackRefs{\CurrentBib}

\bibitem [\protect \citeauthoryear {%
Krueger%
, Rashidi%
\BCBL {}\ \BBA {} Rose%
}{%
Krueger%
\ \protect \BOthers {.}}{%
{\protect \APACyear {2016}}%
}]{%
krueger2016preferences}
\APACinsertmetastar {%
krueger2016preferences}%
\begin{APACrefauthors}%
Krueger, R.%
, Rashidi, T\BPBI H.%
\BCBL {}\ \BBA {} Rose, J\BPBI M.%
\end{APACrefauthors}%
\unskip\
\newblock
\APACrefYearMonthDay{2016}{}{}.
\newblock
{\BBOQ}\APACrefatitle {Preferences for shared autonomous vehicles} {Preferences
  for shared autonomous vehicles}.{\BBCQ}
\newblock
\APACjournalVolNumPages{Transportation Research Part C: Emerging
  Technologies}{69}{}{343--355}.
\PrintBackRefs{\CurrentBib}

\bibitem [\protect \citeauthoryear {%
Lavieri%
\ \BBA {} Bhat%
}{%
Lavieri%
\ \BBA {} Bhat%
}{%
{\protect \APACyear {2019}}%
}]{%
lavieri2019modeling}
\APACinsertmetastar {%
lavieri2019modeling}%
\begin{APACrefauthors}%
Lavieri, P\BPBI S.%
\BCBT {}\ \BBA {} Bhat, C\BPBI R.%
\end{APACrefauthors}%
\unskip\
\newblock
\APACrefYearMonthDay{2019}{}{}.
\newblock
{\BBOQ}\APACrefatitle {Modeling individuals’ willingness to share trips with
  strangers in an autonomous vehicle future} {Modeling individuals’
  willingness to share trips with strangers in an autonomous vehicle
  future}.{\BBCQ}
\newblock
\APACjournalVolNumPages{Transportation Research Part A: Policy and
  Practice}{124}{}{242--261}.
\PrintBackRefs{\CurrentBib}

\bibitem [\protect \citeauthoryear {%
T.~Liu%
, Krishnakumari%
\BCBL {}\ \BBA {} Cats%
}{%
T.~Liu%
\ \protect \BOthers {.}}{%
{\protect \APACyear {2019}}%
}]{%
liu2019canpaxflow}
\APACinsertmetastar {%
liu2019canpaxflow}%
\begin{APACrefauthors}%
Liu, T.%
, Krishnakumari, P.%
\BCBL {}\ \BBA {} Cats, O.%
\end{APACrefauthors}%
\unskip\
\newblock
\APACrefYearMonthDay{2019}{}{}.
\newblock
{\BBOQ}\APACrefatitle {Exploring {D}emand {P}atterns of a {R}ide-{S}ourcing
  {S}ervice using {S}patial and {T}emporal {C}lustering} {Exploring {D}emand
  {P}atterns of a {R}ide-{S}ourcing {S}ervice using {S}patial and {T}emporal
  {C}lustering}.{\BBCQ}
\newblock
\BIn{} \APACrefbtitle {The 6th {IEEE} {I}nternational {C}onference on {M}odels
  and {T}echnologies for {I}ntelligent {T}ransportation {S}ystems.} {The 6th
  {IEEE} {I}nternational {C}onference on {M}odels and {T}echnologies for
  {I}ntelligent {T}ransportation {S}ystems.}
\PrintBackRefs{\CurrentBib}

\bibitem [\protect \citeauthoryear {%
Y.~Liu%
, Bansal%
, Daziano%
\BCBL {}\ \BBA {} Samaranayake%
}{%
Y.~Liu%
\ \protect \BOthers {.}}{%
{\protect \APACyear {2019}}%
}]{%
liu2019framework}
\APACinsertmetastar {%
liu2019framework}%
\begin{APACrefauthors}%
Liu, Y.%
, Bansal, P.%
, Daziano, R.%
\BCBL {}\ \BBA {} Samaranayake, S.%
\end{APACrefauthors}%
\unskip\
\newblock
\APACrefYearMonthDay{2019}{}{}.
\newblock
{\BBOQ}\APACrefatitle {A framework to integrate mode choice in the design of
  mobility-on-demand systems} {A framework to integrate mode choice in the
  design of mobility-on-demand systems}.{\BBCQ}
\newblock
\APACjournalVolNumPages{Transportation Research Part C: Emerging
  Technologies}{105}{}{648--665}.
\PrintBackRefs{\CurrentBib}

\bibitem [\protect \citeauthoryear {%
Ma%
, Zheng%
\BCBL {}\ \BBA {} Wolfson%
}{%
Ma%
\ \protect \BOthers {.}}{%
{\protect \APACyear {2013}}%
}]{%
Ma2013Tshare}
\APACinsertmetastar {%
Ma2013Tshare}%
\begin{APACrefauthors}%
Ma, S.%
, Zheng, Y.%
\BCBL {}\ \BBA {} Wolfson, O.%
\end{APACrefauthors}%
\unskip\
\newblock
\APACrefYearMonthDay{2013}{}{}.
\newblock
{\BBOQ}\APACrefatitle {T-share: A {L}arge-{S}cale {D}ynamic {T}axi
  {R}idesharing {S}ervice} {T-share: A {L}arge-{S}cale {D}ynamic {T}axi
  {R}idesharing {S}ervice}.{\BBCQ}
\newblock
\BIn{} \APACrefbtitle {{IEEE 29th International Conference on Data Engineering
  (ICDE)}} {{IEEE 29th International Conference on Data Engineering (ICDE)}}\
  (\BPGS\ 410--421).
\PrintBackRefs{\CurrentBib}

\bibitem [\protect \citeauthoryear {%
Morales~Sarriera%
\ \protect \BOthers {.}}{%
Morales~Sarriera%
\ \protect \BOthers {.}}{%
{\protect \APACyear {2017}}%
}]{%
morales2017share}
\APACinsertmetastar {%
morales2017share}%
\begin{APACrefauthors}%
Morales~Sarriera, J.%
, Escovar~\'{A}lvarez, G.%
, Blynn, K.%
, Alesbury, A.%
, Scully, T.%
\BCBL {}\ \BBA {} Zhao, J.%
\end{APACrefauthors}%
\unskip\
\newblock
\APACrefYearMonthDay{2017}{}{}.
\newblock
{\BBOQ}\APACrefatitle {To {S}hare or {N}ot {T}o {S}hare: {I}nvestigating the
  {S}ocial {A}spects of {D}ynamic {R}idesharing} {To {S}hare or {N}ot {T}o
  {S}hare: {I}nvestigating the {S}ocial {A}spects of {D}ynamic
  {R}idesharing}.{\BBCQ}
\newblock
\APACjournalVolNumPages{Transportation Research Record: Journal of the
  Transportation Research Board}{2605: 109--117}{}{}.
\PrintBackRefs{\CurrentBib}

\bibitem [\protect \citeauthoryear {%
Qian%
, Zhang%
, Ukkusuri%
\BCBL {}\ \BBA {} Yang%
}{%
Qian%
\ \protect \BOthers {.}}{%
{\protect \APACyear {2017}}%
}]{%
qian2017optimal}
\APACinsertmetastar {%
qian2017optimal}%
\begin{APACrefauthors}%
Qian, X.%
, Zhang, W.%
, Ukkusuri, S\BPBI V.%
\BCBL {}\ \BBA {} Yang, C.%
\end{APACrefauthors}%
\unskip\
\newblock
\APACrefYearMonthDay{2017}{}{}.
\newblock
{\BBOQ}\APACrefatitle {Optimal assignment and incentive design in the taxi
  group ride problem} {Optimal assignment and incentive design in the taxi
  group ride problem}.{\BBCQ}
\newblock
\APACjournalVolNumPages{Transportation Research Part B:
  Methodological}{103}{}{208--226}.
\PrintBackRefs{\CurrentBib}

\bibitem [\protect \citeauthoryear {%
Rayle%
, Dai%
, Chan%
, Cervero%
\BCBL {}\ \BBA {} Shaheen%
}{%
Rayle%
\ \protect \BOthers {.}}{%
{\protect \APACyear {2016}}%
}]{%
rayle2016just}
\APACinsertmetastar {%
rayle2016just}%
\begin{APACrefauthors}%
Rayle, L.%
, Dai, D.%
, Chan, N.%
, Cervero, R.%
\BCBL {}\ \BBA {} Shaheen, S.%
\end{APACrefauthors}%
\unskip\
\newblock
\APACrefYearMonthDay{2016}{}{}.
\newblock
{\BBOQ}\APACrefatitle {Just a better taxi? {A} survey-based comparison of
  taxis, transit, and ridesourcing services in {S}an {F}rancisco} {Just a
  better taxi? {A} survey-based comparison of taxis, transit, and ridesourcing
  services in {S}an {F}rancisco}.{\BBCQ}
\newblock
\APACjournalVolNumPages{Transport Policy}{45}{}{168--178}.
\PrintBackRefs{\CurrentBib}

\bibitem [\protect \citeauthoryear {%
Santi%
\ \protect \BOthers {.}}{%
Santi%
\ \protect \BOthers {.}}{%
{\protect \APACyear {2014}}%
}]{%
Santi2014quantifying}
\APACinsertmetastar {%
Santi2014quantifying}%
\begin{APACrefauthors}%
Santi, P.%
, Resta, G.%
, Szell, M.%
, Sobolevsky, S.%
, Strogatz, S\BPBI H.%
\BCBL {}\ \BBA {} Ratti, C.%
\end{APACrefauthors}%
\unskip\
\newblock
\APACrefYearMonthDay{2014}{}{}.
\newblock
{\BBOQ}\APACrefatitle {Quantifying the benefits of vehicle pooling with
  shareability networks} {Quantifying the benefits of vehicle pooling with
  shareability networks}.{\BBCQ}
\newblock
\APACjournalVolNumPages{Proceedings of the National Academy of Sciences of the
  United States of America}{111: 13290--13294}{}{}.
\newblock
\begin{APACrefDOI} \doi{10.1073/pnas.1403657111} \end{APACrefDOI}
\PrintBackRefs{\CurrentBib}

\bibitem [\protect \citeauthoryear {%
Schneider%
}{%
Schneider%
}{%
{\protect \APACyear {2021}}%
}]{%
todd2019nyc}
\APACinsertmetastar {%
todd2019nyc}%
\begin{APACrefauthors}%
Schneider, T.%
\end{APACrefauthors}%
\unskip\
\newblock
\APACrefYearMonthDay{2021}{}{}.
\newblock
\APACrefbtitle {{NYC} {T}axi \& {R}ide-hailing {S}tats {D}ashboard.} {{NYC}
  {T}axi \& {R}ide-hailing {S}tats {D}ashboard.}
\newblock
\begin{APACrefURL} [{May 27,
  2021}]\url{https://toddwschneider.com/dashboards/nyc-taxi-ridehailing-uber-lyft-data/}
  \end{APACrefURL}
\PrintBackRefs{\CurrentBib}

\bibitem [\protect \citeauthoryear {%
Shaheen%
\ \BBA {} Cohen%
}{%
Shaheen%
\ \BBA {} Cohen%
}{%
{\protect \APACyear {2019}}%
}]{%
shaheen2019shared}
\APACinsertmetastar {%
shaheen2019shared}%
\begin{APACrefauthors}%
Shaheen, S.%
\BCBT {}\ \BBA {} Cohen, A.%
\end{APACrefauthors}%
\unskip\
\newblock
\APACrefYearMonthDay{2019}{}{}.
\newblock
{\BBOQ}\APACrefatitle {Shared ride services in North America: definitions,
  impacts, and the future of pooling} {Shared ride services in north america:
  definitions, impacts, and the future of pooling}.{\BBCQ}
\newblock
\APACjournalVolNumPages{Transport reviews}{39}{4}{427--442}.
\PrintBackRefs{\CurrentBib}

\bibitem [\protect \citeauthoryear {%
Simonetto%
, Monteil%
\BCBL {}\ \BBA {} Gambella%
}{%
Simonetto%
\ \protect \BOthers {.}}{%
{\protect \APACyear {2019}}%
}]{%
simonetto2019real}
\APACinsertmetastar {%
simonetto2019real}%
\begin{APACrefauthors}%
Simonetto, A.%
, Monteil, J.%
\BCBL {}\ \BBA {} Gambella, C.%
\end{APACrefauthors}%
\unskip\
\newblock
\APACrefYearMonthDay{2019}{}{}.
\newblock
{\BBOQ}\APACrefatitle {Real-time city-scale ridesharing via linear assignment
  problems} {Real-time city-scale ridesharing via linear assignment
  problems}.{\BBCQ}
\newblock
\APACjournalVolNumPages{Transportation Research Part C: Emerging
  Technologies}{101}{}{208--232}.
\PrintBackRefs{\CurrentBib}

\bibitem [\protect \citeauthoryear {%
Tachet%
\ \protect \BOthers {.}}{%
Tachet%
\ \protect \BOthers {.}}{%
{\protect \APACyear {2017}}%
}]{%
tachet2017scaling}
\APACinsertmetastar {%
tachet2017scaling}%
\begin{APACrefauthors}%
Tachet, R.%
, Sagarra, O.%
, Santi, P.%
, Resta, G.%
, Szell, M.%
, Strogatz, S\BPBI H.%
\BCBL {}\ \BBA {} Ratti, C.%
\end{APACrefauthors}%
\unskip\
\newblock
\APACrefYearMonthDay{2017}{}{}.
\newblock
{\BBOQ}\APACrefatitle {Scaling {L}aw of {U}rban {R}ide {S}haring} {Scaling
  {L}aw of {U}rban {R}ide {S}haring}.{\BBCQ}
\newblock
\APACjournalVolNumPages{Scientific Reports}{}{}{}.
\newblock
\begin{APACrefDOI} \doi{10.1038/srep42868} \end{APACrefDOI}
\PrintBackRefs{\CurrentBib}

\bibitem [\protect \citeauthoryear {%
Teal%
}{%
Teal%
}{%
{\protect \APACyear {1987}}%
}]{%
Teal1987whohowwhy}
\APACinsertmetastar {%
Teal1987whohowwhy}%
\begin{APACrefauthors}%
Teal, R.%
\end{APACrefauthors}%
\unskip\
\newblock
\APACrefYearMonthDay{1987}{}{}.
\newblock
{\BBOQ}\APACrefatitle {Carpooling: Who, {H}ow and {W}hy} {Carpooling: Who,
  {H}ow and {W}hy}.{\BBCQ}
\newblock
\APACjournalVolNumPages{Transportation Research Part A: General}{21:
  203--214}{}{}.
\newblock
\begin{APACrefDOI} \doi{10.1016/0191-2607(87)90014-8} \end{APACrefDOI}
\PrintBackRefs{\CurrentBib}

\bibitem [\protect \citeauthoryear {%
H.~Wang%
\ \BBA {} Yang%
}{%
H.~Wang%
\ \BBA {} Yang%
}{%
{\protect \APACyear {2019}}%
}]{%
wang2019ridesourcing}
\APACinsertmetastar {%
wang2019ridesourcing}%
\begin{APACrefauthors}%
Wang, H.%
\BCBT {}\ \BBA {} Yang, H.%
\end{APACrefauthors}%
\unskip\
\newblock
\APACrefYearMonthDay{2019}{}{}.
\newblock
{\BBOQ}\APACrefatitle {Ridesourcing systems: A framework and review}
  {Ridesourcing systems: A framework and review}.{\BBCQ}
\newblock
\APACjournalVolNumPages{Transportation Research Part B:
  Methodological}{129}{}{122--155}.
\PrintBackRefs{\CurrentBib}

\bibitem [\protect \citeauthoryear {%
X.~Wang%
, Dessouky%
\BCBL {}\ \BBA {} Ordonez%
}{%
X.~Wang%
\ \protect \BOthers {.}}{%
{\protect \APACyear {2016}}%
}]{%
wang2016pickup}
\APACinsertmetastar {%
wang2016pickup}%
\begin{APACrefauthors}%
Wang, X.%
, Dessouky, M.%
\BCBL {}\ \BBA {} Ordonez, F.%
\end{APACrefauthors}%
\unskip\
\newblock
\APACrefYearMonthDay{2016}{}{}.
\newblock
{\BBOQ}\APACrefatitle {A {P}ickup and {D}elivery {P}roblem for {R}idesharing
  {C}onsidering {C}ongestion} {A {P}ickup and {D}elivery {P}roblem for
  {R}idesharing {C}onsidering {C}ongestion}.{\BBCQ}
\newblock
\APACjournalVolNumPages{Transportation Letters}{8: 259--269}{}{}.
\PrintBackRefs{\CurrentBib}

\bibitem [\protect \citeauthoryear {%
Winter%
, Cats%
, Correia%
\BCBL {}\ \BBA {} van Arem%
}{%
Winter%
\ \protect \BOthers {.}}{%
{\protect \APACyear {2018}}%
}]{%
WINTER2018151}
\APACinsertmetastar {%
WINTER2018151}%
\begin{APACrefauthors}%
Winter, K.%
, Cats, O.%
, Correia, G.%
\BCBL {}\ \BBA {} van Arem, B.%
\end{APACrefauthors}%
\unskip\
\newblock
\APACrefYearMonthDay{2018}{}{}.
\newblock
{\BBOQ}\APACrefatitle {Performance analysis and fleet requirements of automated
  demand-responsive transport systems as an urban public transport service}
  {Performance analysis and fleet requirements of automated demand-responsive
  transport systems as an urban public transport service}.{\BBCQ}
\newblock
\APACjournalVolNumPages{International Journal of Transportation Science and
  Technology}{7: 151-167}{}{}.
\newblock
\begin{APACrefDOI} \doi{https://doi.org/10.1016/j.ijtst.2018.04.004}
  \end{APACrefDOI}
\PrintBackRefs{\CurrentBib}

\bibitem [\protect \citeauthoryear {%
Yap%
, Cats%
\BCBL {}\ \BBA {} van Arem%
}{%
Yap%
\ \protect \BOthers {.}}{%
{\protect \APACyear {2018}}%
}]{%
yap2018crowding}
\APACinsertmetastar {%
yap2018crowding}%
\begin{APACrefauthors}%
Yap, M.%
, Cats, O.%
\BCBL {}\ \BBA {} van Arem, B.%
\end{APACrefauthors}%
\unskip\
\newblock
\APACrefYearMonthDay{2018}{}{}.
\newblock
{\BBOQ}\APACrefatitle {Crowding valuation in urban tram and bus transportation
  based on smart card data} {Crowding valuation in urban tram and bus
  transportation based on smart card data}.{\BBCQ}
\newblock
\APACjournalVolNumPages{Transportmetrica A: Transport Science}{}{}{1-20}.
\PrintBackRefs{\CurrentBib}

\bibitem [\protect \citeauthoryear {%
Zhang%
\ \BBA {} Zhao%
}{%
Zhang%
\ \BBA {} Zhao%
}{%
{\protect \APACyear {2018}}%
}]{%
zhang2018mobility}
\APACinsertmetastar {%
zhang2018mobility}%
\begin{APACrefauthors}%
Zhang, H.%
\BCBT {}\ \BBA {} Zhao, J.%
\end{APACrefauthors}%
\unskip\
\newblock
\APACrefYearMonthDay{2018}{}{}.
\newblock
{\BBOQ}\APACrefatitle {Mobility {S}haring as a {P}reference {M}atching
  {P}roblem} {Mobility {S}haring as a {P}reference {M}atching
  {P}roblem}.{\BBCQ}
\newblock
\APACjournalVolNumPages{IEEE Transactions on Intelligent Transportation
  Systems}{}{}{}.
\PrintBackRefs{\CurrentBib}

\end{thebibliography}
\end{document}